# Jump relations for magnetohydrodynamic shock waves in non-ideal gas flow

R. K. Anand

Department of Physics, University of Allahabad, Allahabad-211 002, India
e-mail: anand.rajkumar@rediffmail.com

**Abstract** The generalized jump relations across the magnetohydrodynamic (MHD) shock front in non-ideal gas are derived considering the equation of state for non-ideal gas as given by Landau and Lifshitz. The jump relations for pressure, density, and particle velocity have been derived, respectively in terms of a compression ratio. Further, the simplified forms of the MHD shock jump relations have been obtained in terms of non-idealness parameter, simultaneously for the two cases viz., (i) when the shock is weak and, (ii) when it is strong. Finally, the cases of strong and weak shocks are explored under two distinct conditions viz., (i) when the applied magnetic field is strong and, (ii) when the field is weak. The aim of this paper is to contribute to the understanding of how shock waves behave in magnetized environment of non-ideal gases.

## 1. Introduction

Shock processes can naturally occur in various astrophysical situations for example, supernova explosions, photoionized gas, stellar winds, collisions between high velocity clumps of interstellar gas, collisions of two or several galaxies etc. Magnetohydrodynamics applies to many conductive fluid and plasma flows encountered in nature and in industrial applications. In numerous circumstances, the flow is subject to a strong as well as a weak magnetic field. This happens in the earth's liquid core and is ubiquitous in solar physics for topics such as sunspots, solar flares, solar corona, solar winds, etc. The strong magnetic fields play significant roles in the dynamics of the interstellar medium. Among the industrial applications involving applied external magnetic fields are drag reduction in duct flows, design of efficient coolant blankets in tokamak fusion reactors, control of turbulence of immersed jets in the steel casting process and advanced propulsion and flow control schemes for hypersonic vehicles.

The magnetic fields have important roles in a variety of astrophysical situations. Complex filamentary structures in molecular clouds, shapes and the shaping of planetary nebulae, synchrotron radiation from supernova remnants, magnetized stellar winds, galactic winds, gamma-ray bursts, dynamo effects in stars, galaxies, and galaxy clusters as well as other interesting problems all involve magnetic fields (see Hartmann, 1998; Balick and Frank, 2002). Due to the formation of filamentary structures in numerical simulations (Porter et al., 1994; Klessen and Burkert, 2000; Ostriker et al., 2001) and the appearance of numerous filaments in observations (Falgarone, Pety and Phillips, 2001), analyses on the processes of filament formation and evolution have also been pursued (Kawachi and Hanawa, 1998; Hennebelle, 2003; Tilley and Pudritz, 2003; Shadmehri, 2005). For these reasons, MHD shock waves are under extensive exploration. Hoffmann and Teller (1950) provided a mathematical treatment for the motion of MHD shock waves in the extreme cases of very weak and very strong magnetic fields. Bazer and Ericson (1959) were among the first to study the hydromagnetic shocks for astrophysical applications. Whang (1984) studied forward-reverse hydromagnetic shock pairs in the Heliosphere. Chakrabarti (1989, 1990) investigated MHD shocks in accretion discs. The physical properties and the evolution of MHD shock waves are significantly modified by the existence of inhomogeneities and boundaries in the stellar atmosphere in which they propagate. Exact analytical solutions giving the evolution of the shock wave were obtained for the case of weak MHD shock waves by FerrizMas and MorenoInsertis (1992). Pogorelov and Matsuda (2000) discussed the nonevolutionary MHD shocks in solar winds. Numerical simulations were carried out to investigate the spherically symmetric shock interaction of pulsar wind nebulae in a supernova remnant (van der Swaluw et al., 2001). Further, Del Zanna, Amato and Bucciantinl (2004) reexamined this problem in an axisymmetric geometry. Ouyed and Pudritz (1993) studied oblique MHD shock waves in disc winds from young stellar objects to explain the broad, blueshifted forbidden emission lines observed in these sources. It is worth mentioning that the magnetic fields play a key role in the formation of the three-ring structure around the supernova 1987A (Tanaka and Washimi, 2002). Takahashi et al. (2002) investigated MHD adiabatic shocks accreting on to a rotating Kerr black hole. Subsequently, Fukumura and Tsuruta (2004) studied the isothermal shock formation around a Keer black hole. Das, Pendharkar and Mitra (2003) studied the isothermal shocks around Schwarzschild black holes using the pseudo-Schwarzschild potential. Koldoba et al. (2008) studied the oscillations of MHD shock waves on the surfaces of T Tauri stars. The analytical solutions were presented by Genot (2009) for anisotropic MHD shocks. Singh, Singh and Husain (2011) obtained the second-kind self-similar solutions to the problem of converging cylindrical shock waves in magnetogasdynamics and recently, Singh (2012) studied the same problem in non-ideal magnetogasdynamics considering the van der Walls equation of state (*see* Wu and Roberts, 1996).

The study of shock waves produced due to the explosions or implosions in presence of magnetic field has received much attention in the past decades and the mainstay of theoretical descriptions is still the Rankine-Hugoniot MHD boundary conditions derived taking into account the equation of state for an ideal gas. Many theoretical and experimental studies were reported by various investigators on planer, cylindrical and spherical MHD shock waves. A number of approaches, including the similarity method (Sedov, 1959; Zel'dovich and Raizer, 1966), power series solution method (Sakurai, 1953, 1954), CCW method (Chester, 1954; Chisnell, 1955; Whitham, 1958), have been used for the theoretical investigations of MHD shock waves in homogeneous and inhomogeneous media. The Rankine-Hugoniot MHD shock conditions (*vide* Appendix) play a fundamental role in all the above well known methods.

The study of MHD shock waves in a non-ideal gas is of great scientific interest in many problems; belong to astrophysics, oceanography, atmospheric sciences, hypersonic aerodynamics and hypervelocity impact. An understanding of the properties of the shock waves in the presence of weak and strong magnetic fields is very useful with regard to the characteristics such as shock strength, shock speed, shock overpressure, and impulse. This has developed potentially my interest in obtaining the generalized MHD shock jump relations for weak and strong shocks in non-ideal gas considering equation of state given by Landau and Lifshitz (1958). Thus, motivated by potentially above wide applications, the main purpose of this paper is to present simplified forms of the useful generalized jump relations for weak and strong MHD shock waves in non-ideal gas, respectively.

In the present paper, the generalized forms of shock jump relations for one-dimensional MHD shock wave propagating in a non-ideal gas are derived which reduce to the well known Rankine-Hugoniot MHD conditions for shock waves in an ideal gas. The equation of state for non-ideal gas is considered as given by Landau and Lifshitz (1958). The shock jump relations for the pressure, density, and particle velocity are obtained, respectively in terms of a single parameter $\xi$, characterizing the shock strength. Besides these jump relations the generalized expression for the velocity of shock propagation is obtained in terms of the same parameter $\xi$. Further, the useful forms of the shock jump relations for pressure, density and particle velocity along with the expression for the shock velocity, in terms of non-idealness parameter are obtained for the two cases: viz., (i) when the shock is weak and (ii) when it is strong, simultaneously. Each case of weak and strong shock is further elaborated under two conditions viz., (i) when the magnetic field is weak and (ii) when the field is strong, respectively. If the non-idealness parameter is taken zero, the generalized MHD shock jump relations reduce to the corresponding jump relations for MHD shock waves in an ideal gas. These generalized MHD shock jump relations for various flow variables are very useful in the theoretical and experimental investigations of strong as well as weak shock waves in the magnetized environments. The paper is structured as follows. The equation of state for non-ideal gas is described in Sect. 2. In Sect. 3, we present the generalized MHD shock conditions. In Sect. 4, we provide results and discussions. In Sect. 5 conclusions are drawn and a brief mention to astrophysical applications is made. The shock conditions corresponding to an ideal gas are summarized in an appendix for the convenience of reference.

**2. The equation of state for non-ideal gas**

A short description of the equation of state for non-ideal gas presented here was given in the recent paper of the author (Anand, 2012). The information given in that paper is repeated here for completeness. The equation of state for a non-ideal gas is obtained by considering an expansion of the pressure $p$ in powers of the density $\rho$ as (*see*, Landau and Lifshitz, 1958)

$$p = \Gamma \rho T [1 + \rho C_1(T) + \rho^2 C_2(T) + .....],$$

where $\Gamma$ is the gas constant, $p$, $\rho$ and $T$ are the pressure, density, and temperature of the non-ideal gas, respectively, and $C_1(T)$, $C_2(T)$, are virial coefficients. The first term in the expansion corresponds to an ideal gas. The second term is obtained by taking into account the interaction between pairs of molecules, and subsequent terms must involve the interactions between the groups of three, four, etc. molecules. In the high temperature range the coefficients $C_1(T)$ and $C_2(T)$ tend to constant values equal to $b$ and $(5/8)b^2$, respectively. For gases $b\rho \ll 1$, $b$ being the internal volume of the molecules, and therefore it is sufficient to consider the equation of state in the form (Anisimov and Spiner, 1972)

$$p = \Gamma \rho T [1 + b\rho] \tag{1}$$

In this equation the correction to pressure is missing due to neglect of second and higher powers of $b\rho$, i.e., due to neglect of interactions between groups of three, four, etc. molecules of the gas. Roberts and Wu (1996, 2003) have used an equivalent equation of state to study the shock theory of sonoluminescence. The internal energy $e$ per unit mass of the non-ideal gas is given as (*see* Anand, 2012)

$$e = p/\rho(\gamma - 1)(1 + b\rho), \tag{2}$$

where $\gamma$ is the adiabatic index. Eq. (2) implies that

$$C_p - C_v = \Gamma(1 + b^2\rho^2/(1 + 2b\rho)) \cong \Gamma$$

neglecting the second and higher powers of $b\rho$. Here $C_p$ and $C_v$ are the specific heats of the gas at constant pressure and constant volume, respectively. Real gas effects can be expressed in the fundamental equations according to Chandrasekhar (1939), by two thermodynamical variables, namely by the sound velocity factor (the isentropic exponent) $\Gamma^*$ and a factor $K$, which contains internal energy as follows,

$$\Gamma^* = (\partial \ln p/\partial \ln \rho)_S \text{ and } K = -\rho(\partial e/\partial \rho)_P / p$$

Using the first law of thermodynamics and the Eqs. (1) and (2), we obtain $\Gamma^* = \gamma(1 + 2b\rho)/(1 + b\rho)$ and $K = 1/(\gamma - 1)$, neglecting the second and higher powers of $b\rho$. This shows that the isentropic exponent $\Gamma^*$ is non-constant in the shocked gas, but the factor $K$ is constant for the simplified equation of state for non-ideal gas in the form given by Eq. (1).

The isentropic velocity of sound, $a$ in non-ideal gas is given by

$$a^2 = \Gamma^* p/\rho \tag{3}$$

### 3. The jump relations across MHD shock front

The work of Hoffmann and Teller (1950), Lundquist (1950), Friedrichs (1955) and Whitham (1958, 1974) is worth mentioning in the context of the present paper. In all of the works, mentioned above, the ambient medium is supposed to be an ideal gas having infinite electrical conductivity. Since at high temperatures that prevail in the problems associated with the shock waves a gas is ionized, electromagnetic effects may also be significant. A complete investigation of such problem should therefore consist of the study of the gasdynamic flow and the electromagnetic fields simultaneously. The fundamental equations of motion, continuity, state and magnetic induction that governs the unsteady, one dimensional motion of MHD shock wave in non-ideal gas, following Whitham (1958), are

$$\frac{\partial \boldsymbol{u}}{\partial t} + (\boldsymbol{u} \cdot \nabla)\boldsymbol{u} + \frac{1}{\rho}\nabla p + \frac{\mu}{\rho}\boldsymbol{H} \times (\nabla \times \boldsymbol{H}) = 0$$

$$\frac{\partial \rho}{\partial t} + \nabla \cdot (\rho \boldsymbol{u}) = 0 \qquad (4)$$

$$\frac{\partial p}{\partial t} + \boldsymbol{u} \cdot \nabla p - \frac{\gamma p(1+2b\rho)}{\rho(1+b\rho)}\left(\frac{\partial \rho}{\partial t} + \boldsymbol{u} \cdot \nabla p\right) = 0$$

$$\frac{\partial \boldsymbol{H}}{\partial t} + \nabla \times (\boldsymbol{H} \times \boldsymbol{u}) = 0$$

where $\boldsymbol{u}$, $p$, and $\rho$ are velocity, pressure, and density of the non-ideal gas behind the shock front, respectively; while $\mu$ is the magnetic permeability and $\boldsymbol{H}$ is the magnetic field. $H^2 = H_r^2 + H_\theta^2 + H_z^2$, where $H_r$, $H_\theta$ and $H_z$ are radial, transverse and axial components of $\boldsymbol{H}$, respectively. In Eq. (4), the viscosity and thermal conductivity are omitted and it is assumed that the fluid has an infinite electrical conductivity so that the diffusion of the magnetic field can be ignored. The Maxwell's equation $\nabla \cdot \boldsymbol{H} = 0$ is included in the first equation of Eq. (4) since the divergence of that equation gives $\partial(\nabla \cdot \boldsymbol{H})/\partial t = 0$. The propagation of shock waves under the influence of weak and strong magnetic fields constitutes a problem of great interest to researches in many branches of science and astrophysics. In the investigation of such problems it is assumed that $\boldsymbol{u}$ is radial and that all flow quantities are functions of the radial distance $r$ and the time $t$. The transverse and axial components $H_\theta$, $H_z$ of $\boldsymbol{H}$ need not vanish, however. Indeed they are of primary importance since it is easily shown that the radial component $H_r$ must vanish or the solution becomes trivial. For, if $H_r \neq 0$, the condition that $\theta$ and $z$ components of $\boldsymbol{H} \times (\nabla \times \boldsymbol{H})$ must vanish in the third equation of Eq. (4) requires that $rH_\theta$ and $H_z$ be independent of $r$. In turn these lead to artificial forms for $u$ and the only feasible case turns out to be $H_\theta = H_z = 0$. But then the flow problem is independent of the magnetic field. Whitham (1958), Sakurai (1962), Summer (1975), Philip and Shimshon (1976), and many others have studied shock waves through ideal gas atmosphere in the presence of magnetic field taking into account only either axial component or transverse component or both transverse and axial components.

The jump conditions across shock front relate the fluid properties behind the shock, which is referred to as the downstream region, to the fluid properties ahead of the shock, which is referred to as the upstream region. These conditions are derived from the principles of conservation of mass, magnetic flux, momentum and energy under the assumption that the shock is a single unsteady wave front with no thickness or a single steady wave of finite thickness. If a planar shock front exists, then the velocity vector $u$ and the magnetic field vector $H$ can both be broken down into components perpendicular to the shock front and parallel to the shock front. The most mathematically tractable MHD shock is a normal shock (parallel component of $u = 0$, and perpendicular component of $u = u$) in which the magnetic field is parallel to the shock front (i.e., perpendicular component of $H = 0$, and parallel component of $H = H$). In a frame of reference moving with the shock front, the jump conditions at the MHD shock in non-ideal gas flow are given by the principles of conservation of mass, magnetic flux, momentum and energy across the shock, namely,

$$\rho(U - u) = \rho_o U$$

$$H(U - u) = H_o U$$

$$p + \frac{1}{2}\mu H^2 + \rho(U-u)^2 = p_o + \frac{1}{2}\mu H_o^2 + \rho_o U^2 \qquad (5)$$

$$\frac{1}{2}(U-u)^2 + \frac{p}{\rho}\left[\frac{1+(\gamma-1)(1+b\rho)}{(\gamma-1)(1+b\rho)}\right] + \frac{\mu H^2}{\rho} = \frac{1}{2}U^2 + \frac{p_o}{\rho_o}\left[\frac{1+(\gamma-1)(1+b\rho_o)}{(\gamma-1)(1+b\rho_o)}\right] + \frac{\mu H_o^2}{\rho_o}$$

where $U, u, \rho, H$ and $\mu$ are, respectively, the shock velocity, particle velocity, density, magnetic field and magnetic permeability of the non-ideal gas. $H_o^2 = H_{ro}^2 + H_{\theta o}^2 + H_{zo}^2$, where $H_{ro}, H_{\theta o}$ and $H_{zo}$ are radial, transverse and axial components of $H_o$, respectively. The quantities with the suffix $o$ and without suffix denote the values of the quantities in upstream region i.e., immediately ahead of shock, and in downstream region i.e., immediately behind the shock, respectively. Also, the effects of viscosity and thermal conductivity are omitted in Eq. (5), and it is assumed that the non-ideal gas has an infinite electrical conductivity. It is worth mentioning that Eq. (5) is also valid for curved shock fronts (e.g., in a spherical medium) because the thickness of the front is almost always negligible compared to its radius of curvature. The Eq. (5) is utilized to derive the generalized shock jump relation for one-dimensional, MHD shock waves in non-ideal gas. A comprehensive set of explicit formulae relating downstream and upstream variables can be conveniently derived in terms of a single parameter $\xi(=\rho/\rho_o)$, characterizing the shock strength, as

$$\rho = \rho_o \xi, \quad H = H_o \xi, \quad u = U(\xi - 1)/\xi,$$

$$p = p_o + \frac{2\rho_o(\xi-1)}{(\gamma+1)-(\gamma-1)\xi+b\rho_o\{2+5\gamma-\xi(2\gamma-3)-\xi^2(\gamma-1)\}} \times$$
$$\left[ a_o^2\{1+b\rho_o(1+\xi)\} + \frac{b_o^2}{4}\{(\gamma-1)(\xi-1)^2 + b\rho_o\{(\gamma-1)(2+\xi^3)+\xi(3\gamma-2)(2\xi-3)\}\} \right] \quad (6)$$

$$U^2 = \frac{2\xi}{(\gamma+1)-(\gamma-1)\xi+b\rho_o\{2+5\gamma-\xi(2\gamma-3)-\xi^2(\gamma-1)\}} \times$$
$$\left[ a_o^2\{1+b\rho_o(1+\xi)\} + b_o^2\left\{\left(1-\frac{\gamma}{2}\right)\xi + \frac{\gamma}{2} + b\rho_o(3\xi+\gamma)\right\} \right]$$

where $a_o\left(=\sqrt{\gamma \delta p_o/\rho_o}\right)$ is speed of sound in medium ahead of the shock front, $\delta = (1+2b\rho_o)/(1+b\rho_o)$, and $b\rho_o$ is the parameter of non-idealness, and $b_o\left(=\sqrt{\mu H_o^2/\rho_o}\right)$, is the Alfven speed (i.e., speed of magnetic waves in an incompressible fluids).

**3.1 Jump relations for weak shocks:** For very weak shock, the parameter $\xi$ can be written as $\xi = 1+\varepsilon$, where $\varepsilon$ is another parameter which is negligible in comparison with unity, i.e., $\varepsilon \ll 1$.

Now, let us consider weak shocks in weak and strong magnetic fields, respectively, as

3.1.1 For weak shock in weak magnetic field (WSWMF), $b_o^2 \ll a_o^2$ i.e., $\mu H_o^2/\gamma p_o \ll 1$, under this condition the generalized jump relations (6) reduce to

$$\rho = \rho_o(1+\varepsilon), \; H = H_o(1+\varepsilon), \; p = p_o[1+\gamma(1-b\rho_o\gamma)\varepsilon], \quad (7)$$
$$u = a_o(1-b\rho_o\gamma/4 - 5b\rho_o/4)\varepsilon, \; U = [1-b\rho_o(5+\gamma)/4 + \{1+\gamma-2b\rho_o(4-\gamma)\}\varepsilon/4]a_o$$

It should be noted that in this limiting case, the expressions for $p$, $\rho$, $U$ and $u$ in terms of $\varepsilon$ are independent of the magnetic field.

3.1.2 For weak shock in strong magnetic field (WSSMF), $b_o^2 \gg a_o^2$ i.e., $\mu H_o^2/\gamma p_o \gg 1$, under this condition the generalized jump relations (6) reduce to

$$\rho = \rho_o(1+\varepsilon),\ H = H_o(1+\varepsilon),\ p = p_o\left[1+\{\gamma - b\rho_o b_o^2/4\}\varepsilon\right],\ u = \varepsilon b_o, \quad (8)$$

$$U = \left[1 + \{3/4 + b\rho_o(\gamma - 2 - \gamma^2/4)\}\varepsilon\right]b_o$$

**3.2 Jump relations for strong shocks:** There are two possibilities leading to strong shocks, i.e., large values of $p/p_o$; either $\xi$ is close to the value of $(\gamma+1)/(\gamma-1)$ or $b_o^2/a_o^2$ is large. The former is the usual case found in gas dynamics, but it is an interesting fact that in magnetohydrodynamics strong shocks may also arise when the magnetic field is very strong for any value of $\xi > 1$. Now let us consider the two cases of strong and weak magnetic fields:

3.2.1 For strong shock in weak magnetic field (SSWMF), $b_o^2 \ll a_o^2$ i.e., $\mu H_o^2/\gamma p_o \ll 1$, under this condition the generalized jump relations (6) reduce to

$$\rho = \rho_o \xi,\ H = H_o \xi,\ u = U(\xi-1)/\xi,\ p/p_o = 1 + \delta\{\chi' a_o^2 + A' b_o^2\}U^2/a_o^4 \quad (9)$$

where $\chi' = \dfrac{\gamma(\xi-1)}{\xi}$,

$$A' = \frac{\gamma(\xi-1)}{4\xi\{1+b\rho_o(1+\xi)\}}\left[\begin{array}{l}(\gamma-1)(\xi-1)^2 - 2\{(2-\gamma)\xi + \gamma\} + \\ + b\rho_o\{(2+\xi^3)(\gamma-1) + \xi(3\gamma-2)(3\xi-2) - 2(3\xi+\gamma)\}\end{array}\right]$$

3.2.2 For strong shock in strong magnetic field (SSSMF), $b_o^2 \gg a_o^2$ i.e., $\mu H_o^2/\gamma p_o \gg 1$, under this condition the generalized jump relations (6) reduce to

$$\rho = \rho_o \xi,\ H = H_o \xi,\ u = U(\xi-1)/\xi,$$

$$p/p_o = 1 + \chi\{b_o^2 + A[1+b\rho_o(1+\xi)]a_o^2\}U^2/a_o^2 b_o^2 \quad (10)$$

where $\chi = \delta\dfrac{\gamma(\xi-1)\{(\gamma-1)(\xi-1)^2 + b\rho_o\{(\gamma-1)(2+\xi^3) + \xi(3\gamma-2)(2\xi-3)\}\}}{2\xi\{(2-\gamma)\xi + \gamma + 2b\rho_o(3\xi+\gamma)\}}$

$$A = \frac{4}{(\gamma-1)(\xi-1)^2 + b\rho_o\{(2+\xi^3)(\gamma-1) + \xi(3\gamma-2)(2\xi-3)\}} - \frac{2}{(2-\gamma)\xi + \gamma + 2b\rho_o(3\xi+\gamma)}$$

## 4. Results and discussions

The generalized forms of shock jump relations for one-dimensional MHD shock waves propagating in non-ideal gas are derived which reduce to the Rankine-Hugoniot MHD conditions for an ideal gas when non-idealness parameter becomes zero. The shock jump relations for the pressure, density, and particle velocity are obtained, respectively in terms of a single parameter $\xi$, characterizing the shock strength. Besides these jump relations the generalized expression for the shock velocity is also obtained in terms of the same parameter $\xi$. These relations are given by Eq. (6). Further, the useful forms of shock jump relations for the pressure, density and particle velocity and also the expression for the shock velocity, in terms of non-idealness parameter are obtained for the two cases: viz., (i) when the shock is weak and (ii) when it is strong, simultaneously. Each case of weak and strong shock is further elaborated under two conditions viz., (i) when the magnetic field is weak and (ii) when the field is strong, respectively. The useful forms of the generalized MHD shock jump relations are valid for the exploding and imploding shock waves.

The generalized expressions for the shock velocity $U/a_o$, pressure $p/p_o$, density $\rho/\rho_o$, and particle velocity $u/a_o$ immediately behind the MHD shock front are given by Eq. (6). The strength of magnetic field is measured by the ratio of Alfven speed to sound speed in medium ahead of the shock front i.e., $b_o^2/a_o^2 = \beta^2$ (say). For the purpose of numerical calculations, values of the non-idealness parameter are taken to be 0 (for ideal gas), 0.03125, 0.06250, 0.09375, 0.12500, 0.15625, 0.18750, 0.21875 and 0.25000. The variations of shock velocity, pressure and particle velocity with $\xi$ for $\gamma = 1.4$, $\beta^2 = 10$ and different values of non-idealness parameter $b\rho_o$ are shown through Fig. 1. It is remarkable that the shock velocity, pressure and particle velocity increase rapidly with the shock strength $\xi$ (especially when $\xi \geq 4$) as well as with an increase in the value of non-idealness parameter. Fig.1. reveals clearly the influence of magnetic field that in the presence of magnetic field there is large increase in the values of flow variables behind the shock especially in the case of pressure whereas in the absence of magnetic field there is small increase in the values of flow variables. Also, in the absence of magnetic field the shock velocity remains unchanged with the shock strength $\xi$, whereas the pressure and particle velocity become constant when $\xi \geq 3$, for $b\rho_o = 0$. Fig. 2 shows the variations of shock velocity, pressure and particle velocity with $\beta^2$ for $\gamma = 1.4$, $\xi = 3$ and different values of non-idealness parameter $b\rho_o$. It is notable that the shock velocity, pressure and particle velocity increase with the strength of magnetic field, and also with an increase in the value of non-idealness parameter. Obviously, in the presence of magnetic field the shock propagates more rapidly in real gases comparatively in an ideal gas. Also, the pressure, and particle velocity behind the shock front increase quickly in the presence of magnetic field. It is interesting to note that the rate of rise in the flow variables increases with increase in the strength of magnetic field.

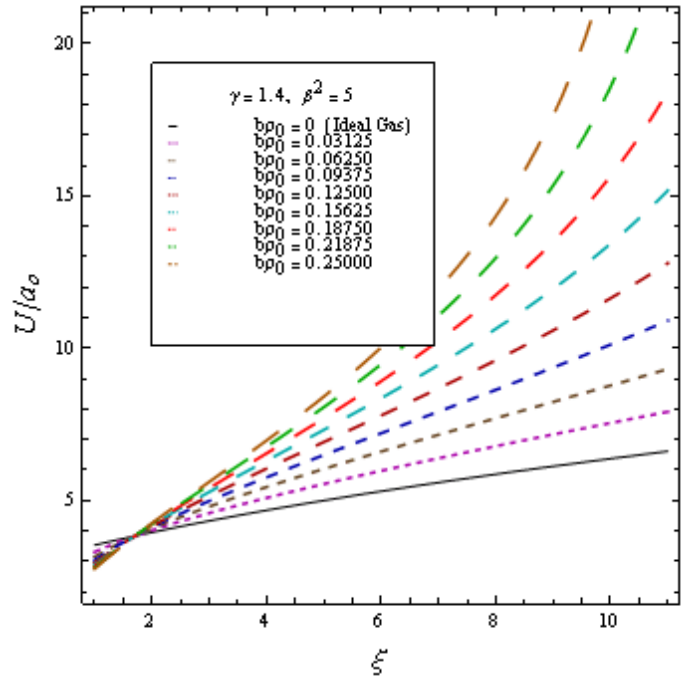

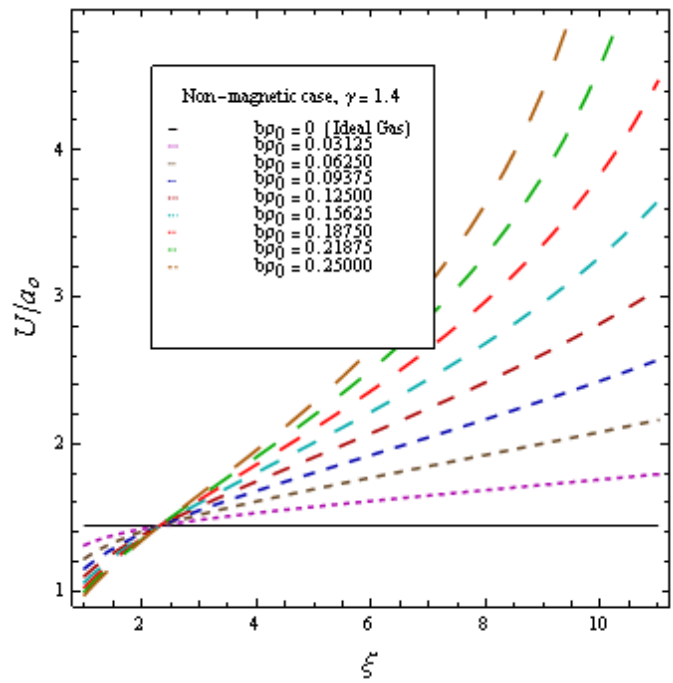

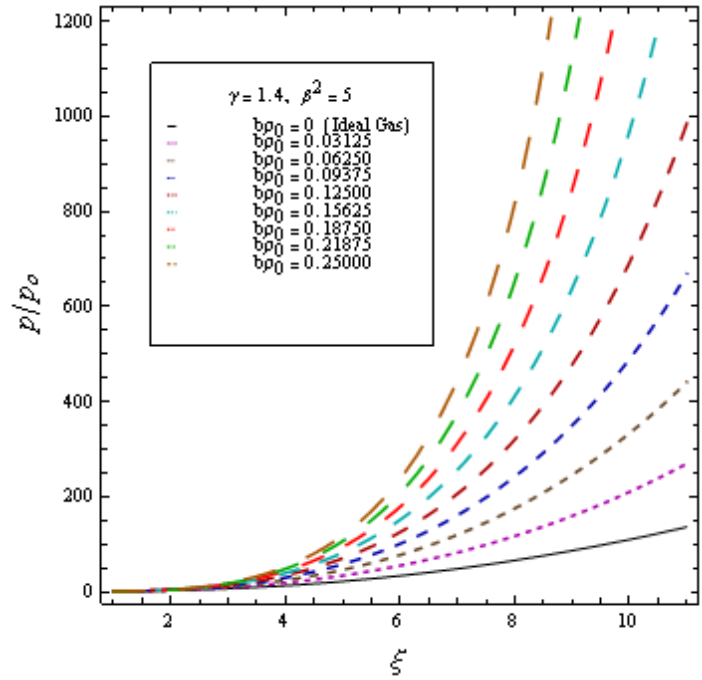

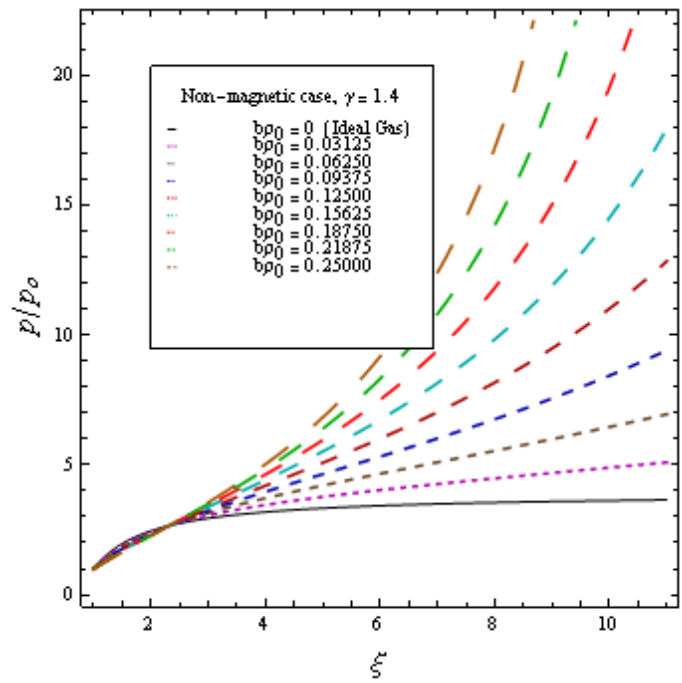

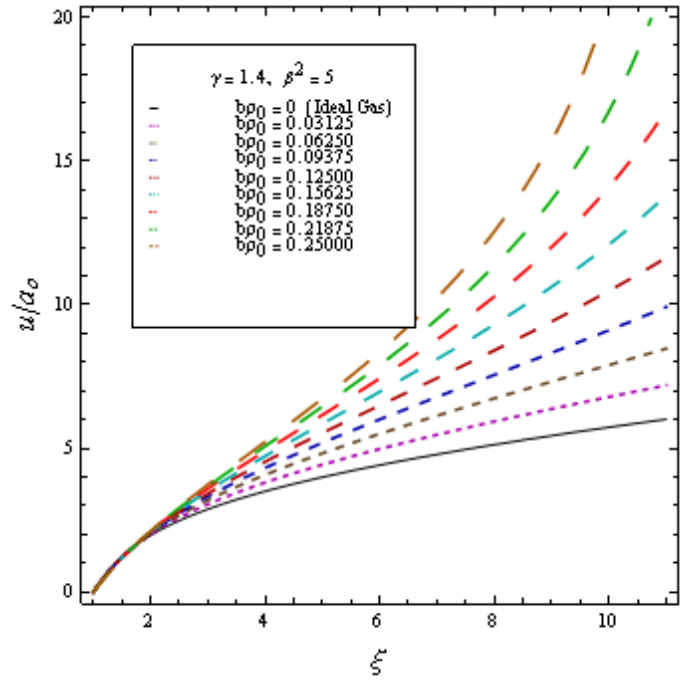

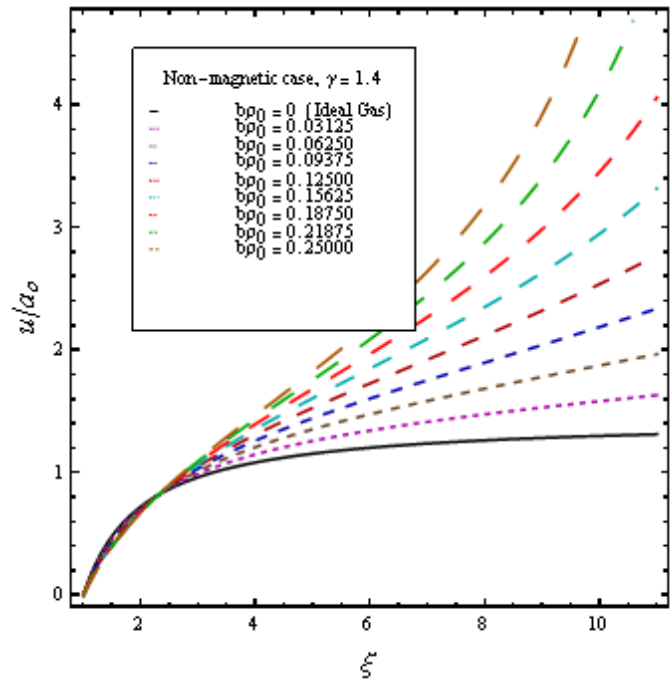

**Fig.1** Variations of $U/a_o$, $p/p_o$ and $u/a_o$ with $\xi$ for different values of $b\rho_o$

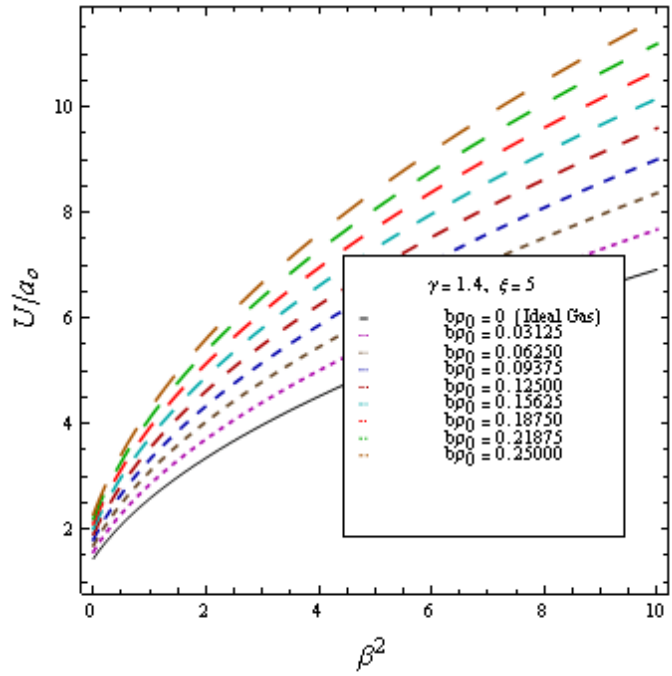

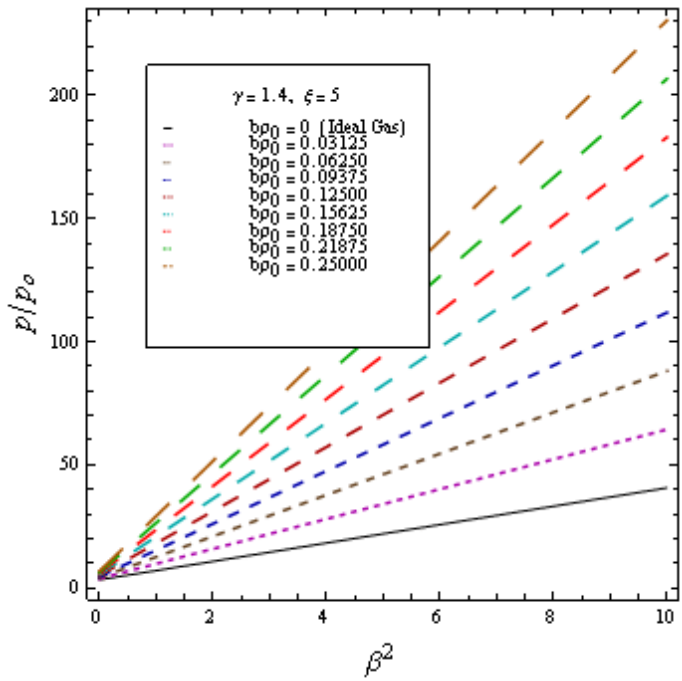

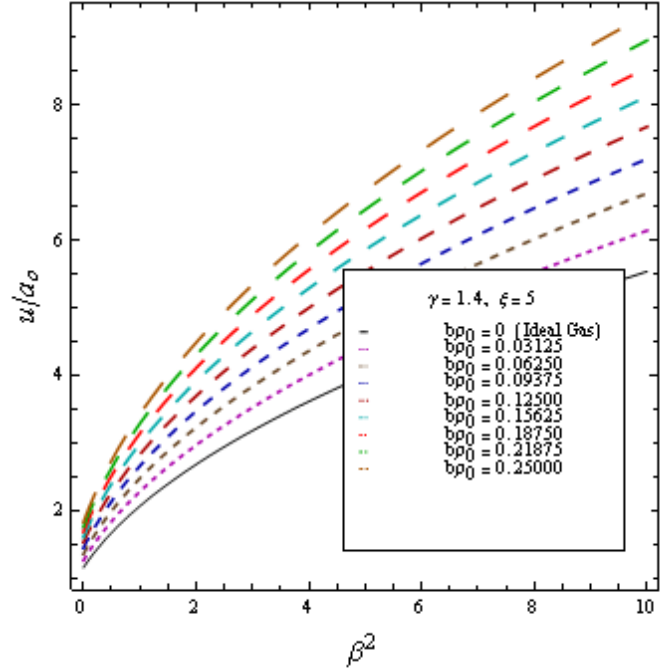

**Fig.2** Variations of $U/a_o$, $p/p_o$ and $u/a_o$ with $\beta^2$ for different values of $b\rho_o$

**4.1 Weak shock waves** Now we explore the influence of magnetic field on the weak shock waves propagating in non-ideal gas under two conditions viz., (i) when the magnetic field is weak and (ii) when the field is strong, respectively.

4.1.1 Weak shock in weak magnetic field: The generalized MHD shock jump relations for weak shocks in weak magnetic field are given by Eq. (7). These relations i.e., shock velocity $U/a_o$, pressure $p/p_o$, and particle velocity $u/a_o$ are dependent of $\varepsilon(r)$ a parameter which is negligible in comparison with unity, adiabatic index $\gamma$, strength of magnetic field $\beta^2$, and non-idealness parameter $b\rho_o$. The variations in the shock velocity, pressure, and particle velocity with $\varepsilon$ for $\gamma = 1.4$ and different values of non-idealness parameter $b\rho_o$, are shown in Fig. 3. It is important to note that the shock velocity, pressure, and particle velocity increase with increase in the parameter $\varepsilon$. It is also seen that an increase in the value of non-idealness parameter leads to a decrease in the shock velocity, pressure and particle velocity. It is worth mentioning that the shock velocity, pressure, and particle velocity are independent of the strength of magnetic field $\beta^2$. Thus, weak magnetic field does not influence the properties of weak shock waves. Obviously, the trends of variations of flow quantities behind the shock front in real gases are similar to that of behind the shock front in an ideal gas.

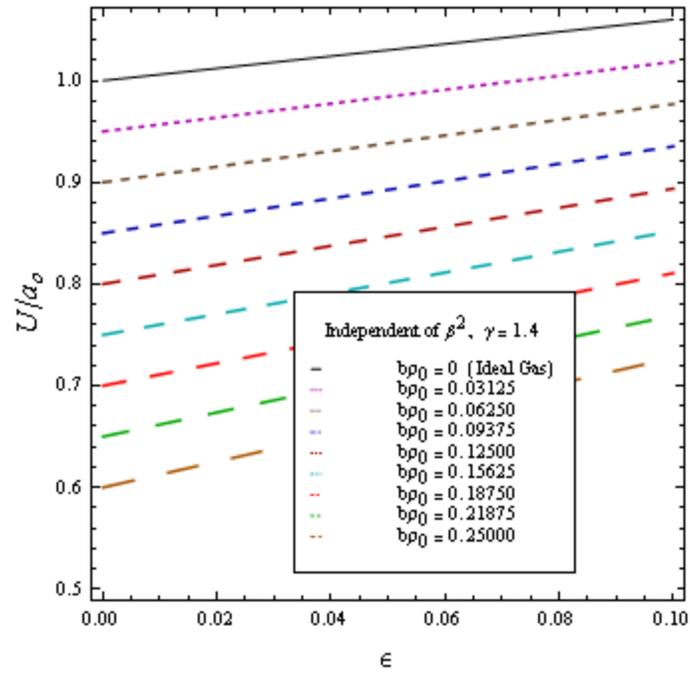

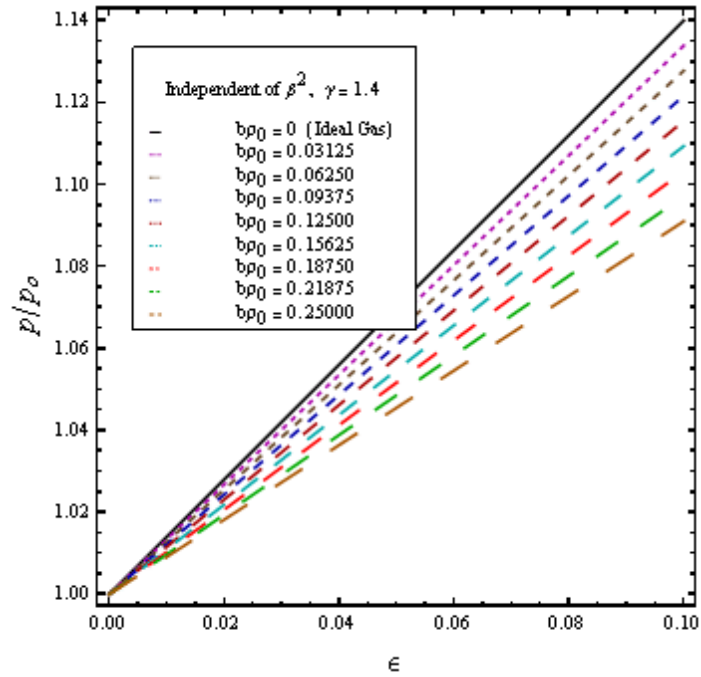

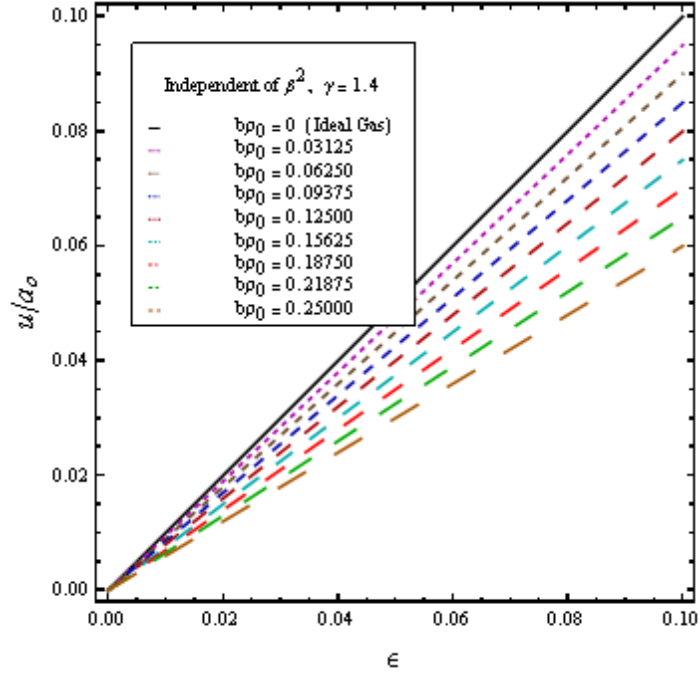

**Fig.3** Variations of $U/a_o$, $p/p_o$ and $u/a_o$ with $\varepsilon$ for different values of $b\rho_o$

4.1.2 Weak shock in strong magnetic field: The generalized MHD shock jump relations for weak shocks in strong magnetic field are given by Eq. (8). These relations i.e., shock velocity $U/a_o$, pressure $p/p_o$, and particle velocity $u/a_o$ are dependent of $\varepsilon(r)$ a parameter which is negligible in comparison with unity, adiabatic index $\gamma$, strength of magnetic field $\beta^2$, and non-idealness parameter $b\rho_o$. The variations in shock velocity, pressure, and particle velocity with $\varepsilon$ for $\gamma = 1.4$, $\beta^2 = 5$ and different values of non-idealness parameter $b\rho_o$, are shown in Fig. 4. It is found that the shock velocity, pressure, and particle velocity increase with increase in the parameter $\varepsilon$. It is also seen that an increase in the value of non-idealness parameter leads to a decrease in the shock velocity and pressure. However, the particle velocity remains unchanged with the non-idealness parameter. In the absence of magnetic field, the shock velocity and particle velocity remain unchanged whereas the pressure increases with increase in the parameter $\varepsilon$. The variations in shock velocity, pressure, and particle velocity with $\beta^2$ for $\gamma = 1.4$, $\varepsilon = 0.05$ and different values of non-idealness parameter $b\rho_o$, are shown in Fig. 5. The shock velocity and particle velocity increase whereas the pressure except for $b\rho_o = 0$ decreases with increase in the strength of magnetic field. However, the pressure for $b\rho_o = 0$ (Ideal gas) remains constant with the strength of magnetic field. It is also seen that an increase in the value of non-idealness parameter leads to a decrease in the pressure. It is interesting to note that the particle velocity is independent of the non-idealness parameter, and adiabatic index. Also, it is worth mentioning that the particle velocity remains unchanged with the parameter $\varepsilon$, whereas it increases with the strength of magnetic field. It is worth mentioning that the trends of variations of flow quantities behind the shock front in real gases are similar to that of behind the shock front in an ideal gas.

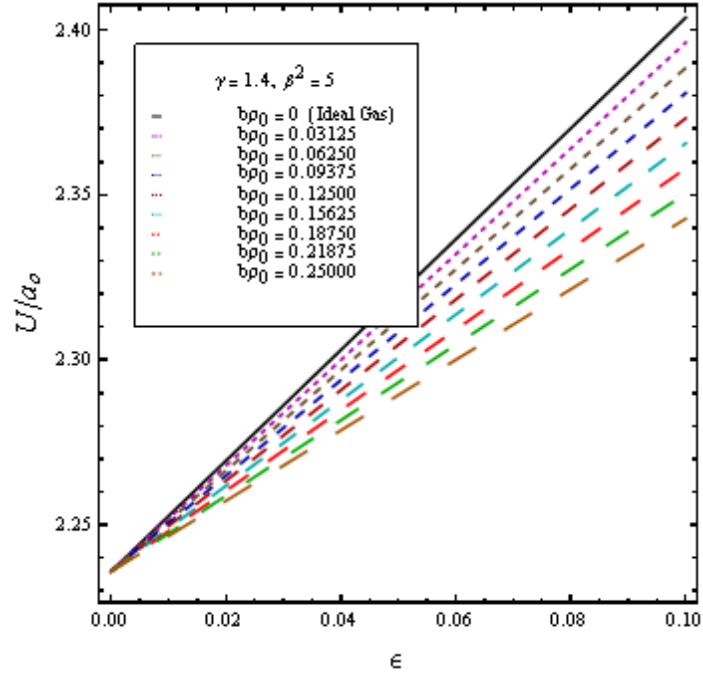

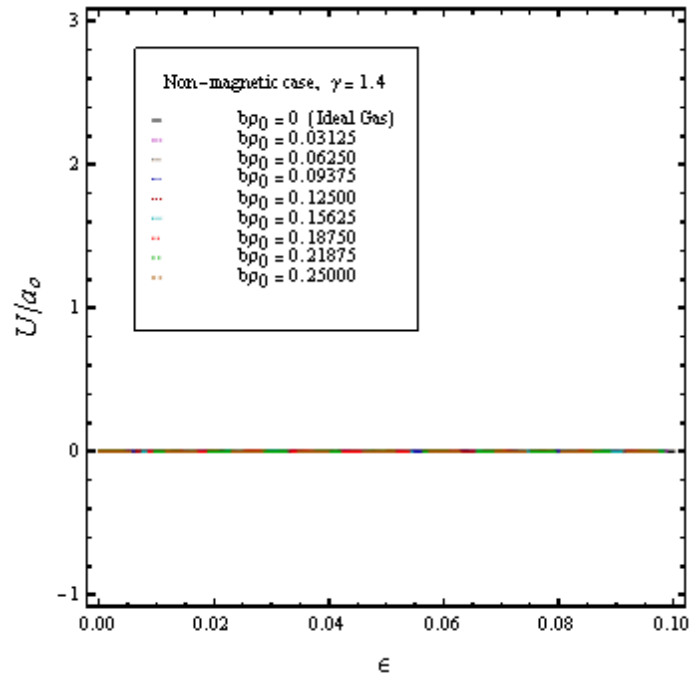

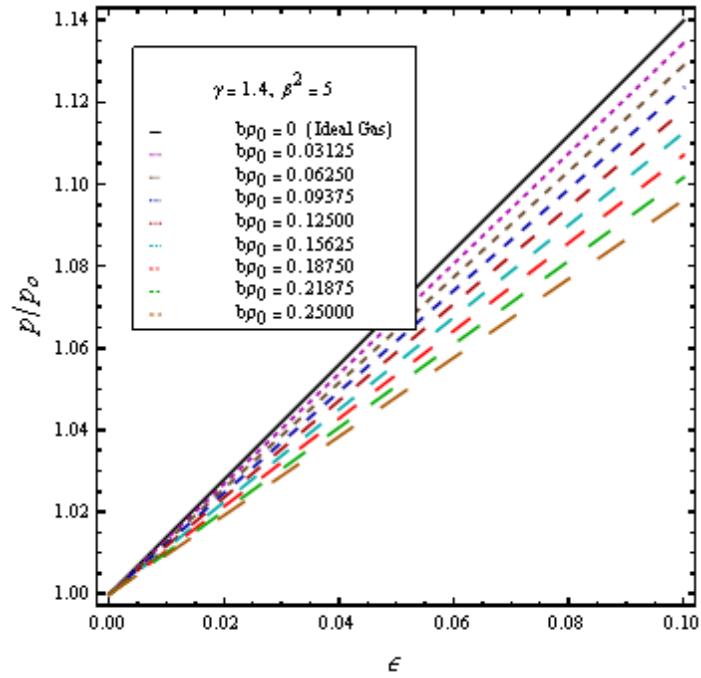

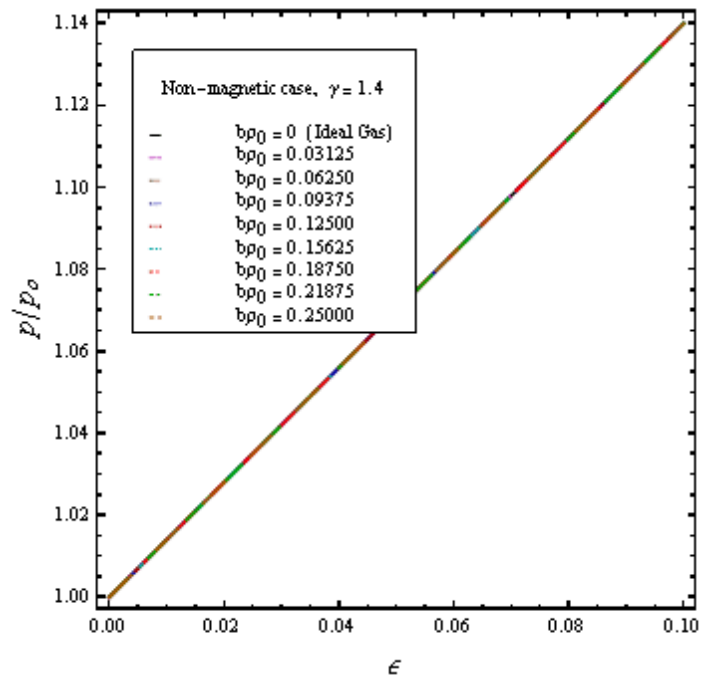

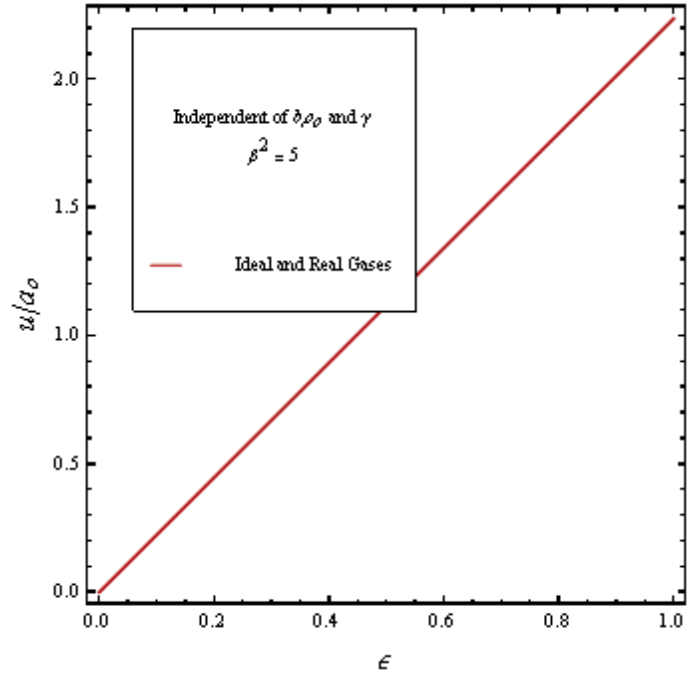

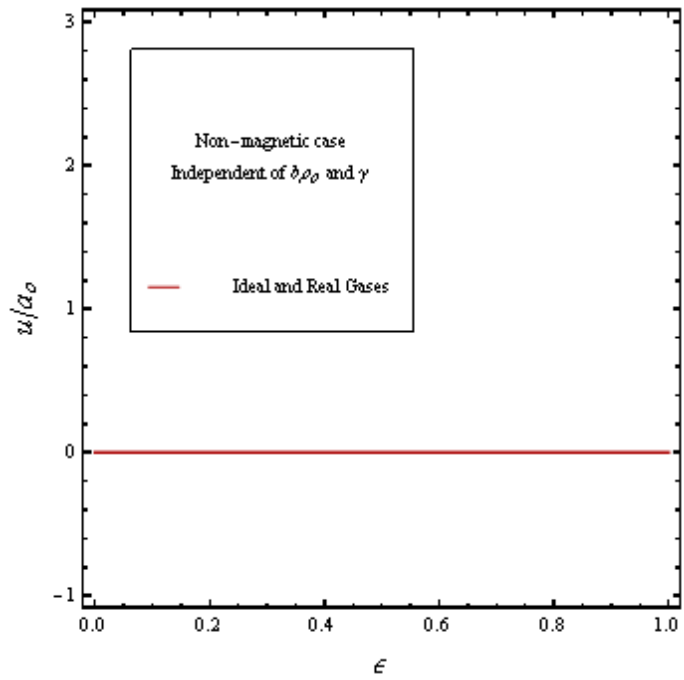

**Fig. 4** Variations of $U/a_o$, $p/p_o$ and $u/a_o$ with $\varepsilon$ for different values of $b\rho_o$

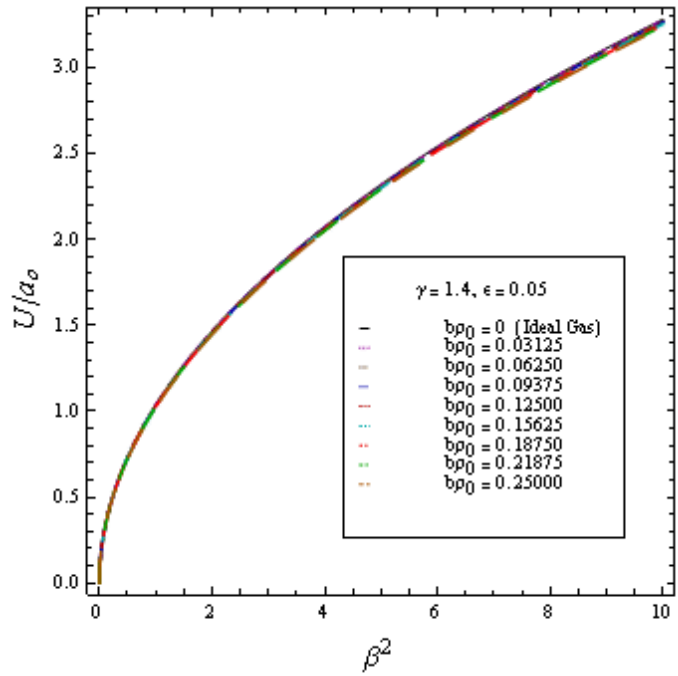

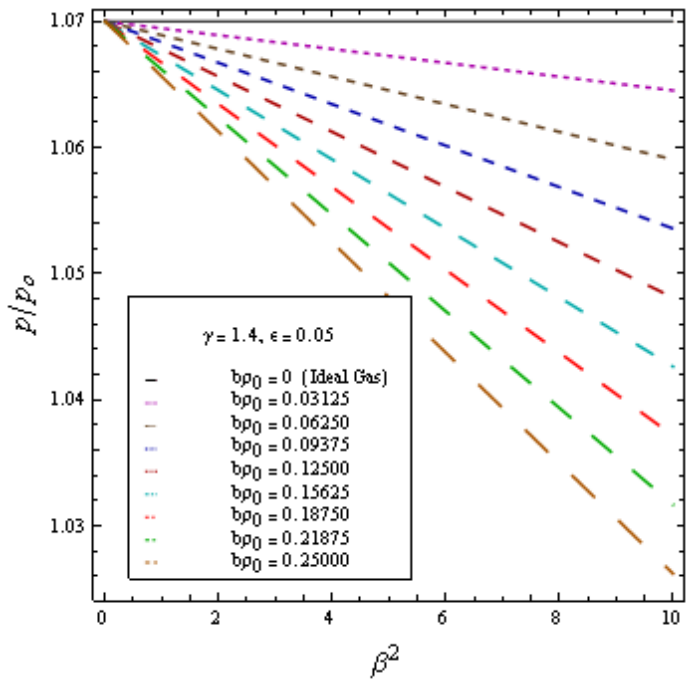

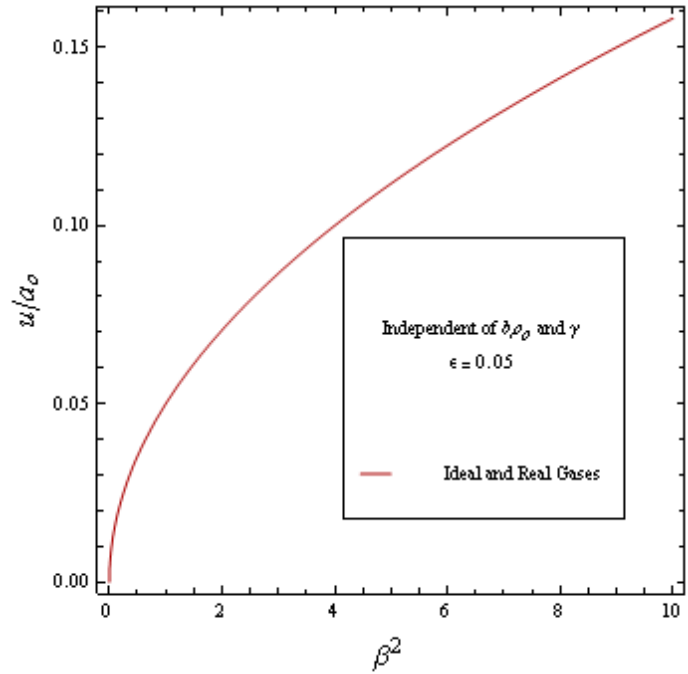

**Fig. 5** Variations of $U/a_o$, $p/p_o$ and $u/a_o$ with $\beta^2$ for different values of $b\rho_o$

**4.2 Strong shock waves** Now we explore the influence of magnetic field on the strong shock waves propagating in non-ideal gas under two conditions viz., (i) when the magnetic field is weak and (ii) when the field is strong, respectively.

4.2.1 Strong shock in weak magnetic field: The generalized MHD shock jump relations for strong shocks in weak magnetic field are given by Eq. (9). These relations i.e., pressure $p/p_o$ and particle velocity $u/a_o$, are dependent of $\xi$ a parameter characterizing the shock strength, strength of magnetic field $\beta^2$, adiabatic index $\gamma$, non-idealness parameter $b\rho_o$ and shock velocity $U/a_o$. The variations in the pressure, speed of sound and particle velocity with $\xi$ for $\gamma = 1.4$, $\beta^2 = 0.5$, $U/a_o = 3$ and different values of non-idealness parameter $b\rho_o$, are shown in Fig. 6. It is important to note that the pressure, speed of sound, and particle velocity increase with increase in the shock strength. It is also seen that an increase in the value of non-idealness parameter leads to an increase in the pressure and speed of sound. It is worth mentioning that the particle velocity does not depend on the non-idealness parameter, and adiabatic index. It is worth mentioning that in the absence of magnetic field both the pressure and speed of sound first increase rapidly with increase in the shock strength and then the pressure becomes nearly constant however the speed of sound achieves a maximum value and then starts to decrease with the shock strength. Fig. 7 shows the variations of pressure, speed of sound and particle velocity with $\beta^2$ for $\gamma = 1.4$, $\xi = 5$, $U/a_o = 3$ and different values of non-idealness parameter $b\rho_o$. It is important to note that the pressure and speed of sound increase (except for $b\rho_o = 0$) with increase in the strength of magnetic field, whereas the pressure and speed of sound decrease for $b\rho_o = 0$. It is also seen that an increase in the value of non-idealness parameter leads to an increase in the pressure, and speed of sound. The particle velocity remains unchanged with the strength of magnetic field. The variations in the pressure, speed of sound and particle velocity with $U/a_o$ for $\gamma = 1.4$, $\xi = 5$, $\beta^2 = 0.5$ and different values of non-idealness parameter $b\rho_o$, are shown in Fig. 8. It is important to note that the pressure, speed of sound and particle velocity increase with increase in the shock velocity. It is also seen that an increase in the value of non-idealness parameter leads to an increase in the pressure, and particle velocity. It is worth mentioning that the particle velocity increases rapidly with increase in the shock strength, and it increases linearly with increase in the shock velocity, in both the magnetic and non-magnetic cases. Also, the particle velocity is independent of the non-idealness parameter, and adiabatic index. Obviously, the trends of variations of flow quantities behind the shock front in real gases are similar to that of behind the shock front in an ideal gas. It is notable that the pressure and speed of sound increase with the non-idealness parameter whereas the particle velocity is independent of the non-idealness parameter.

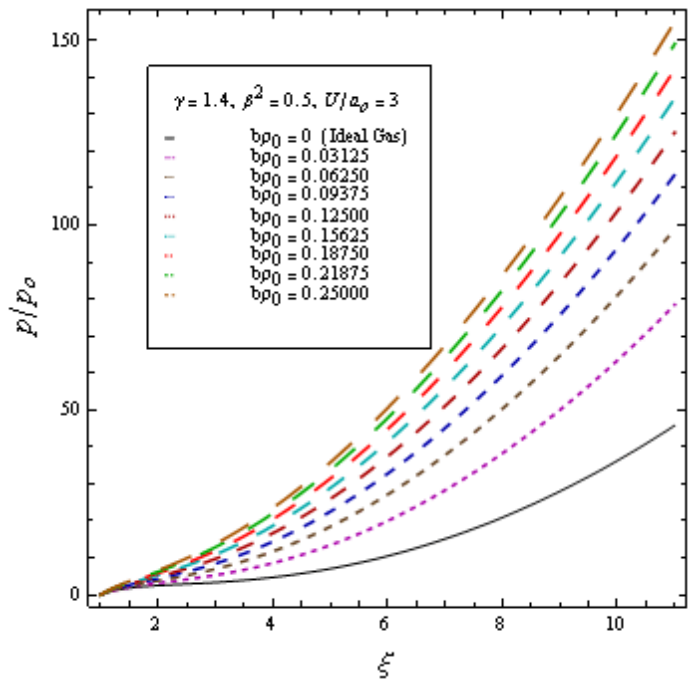

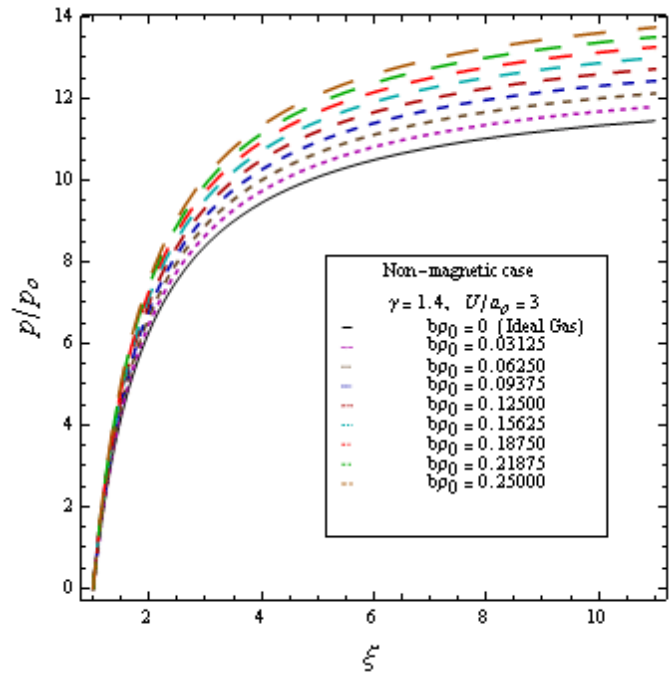

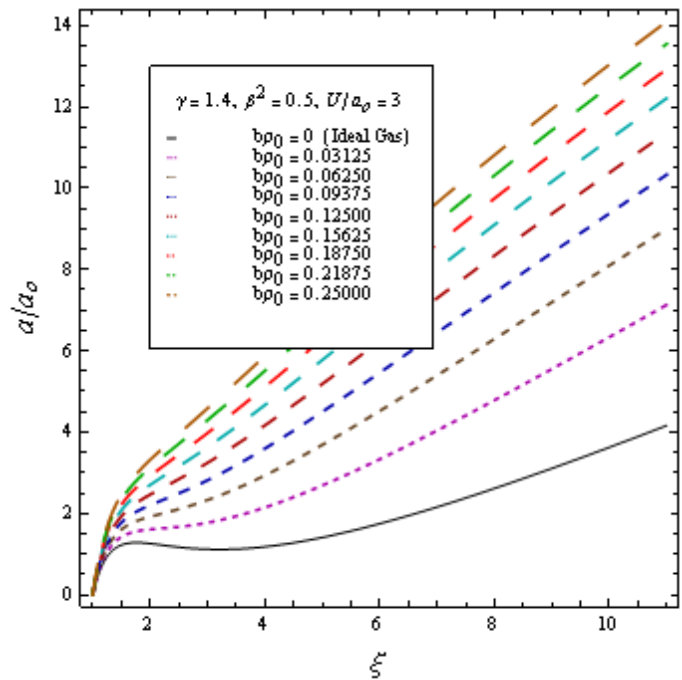

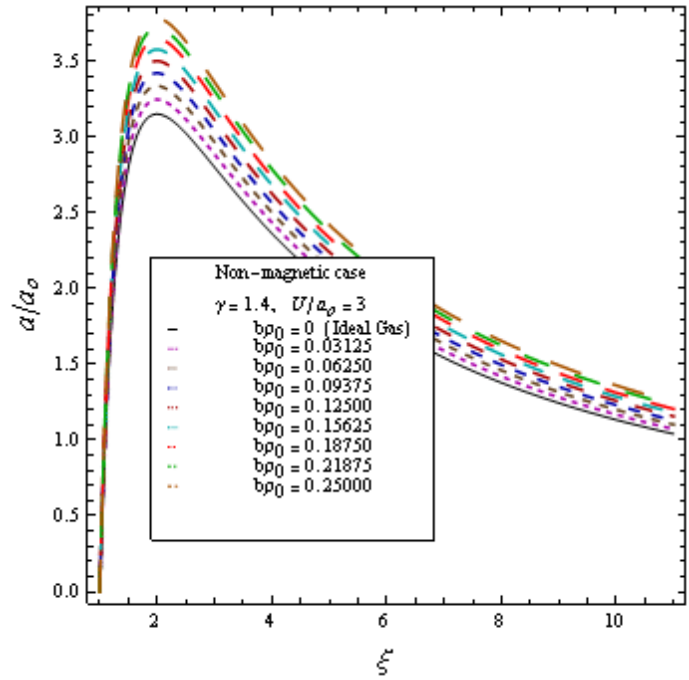

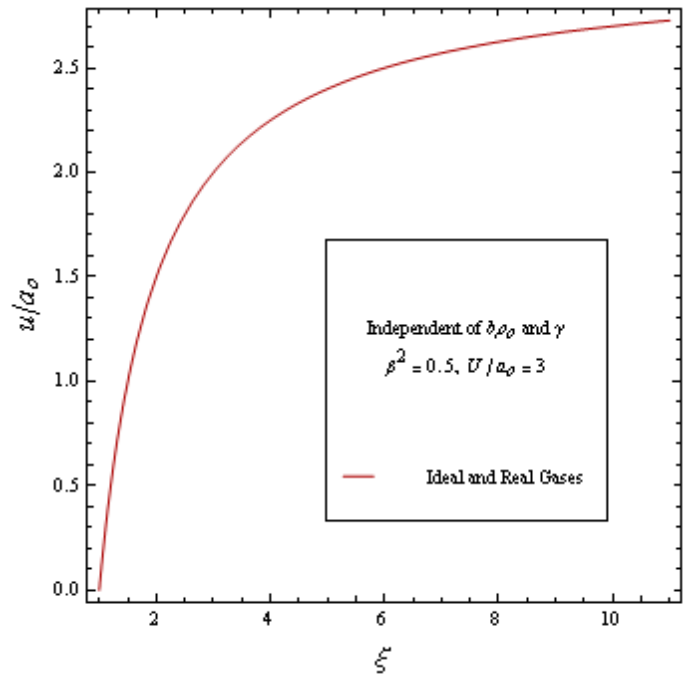

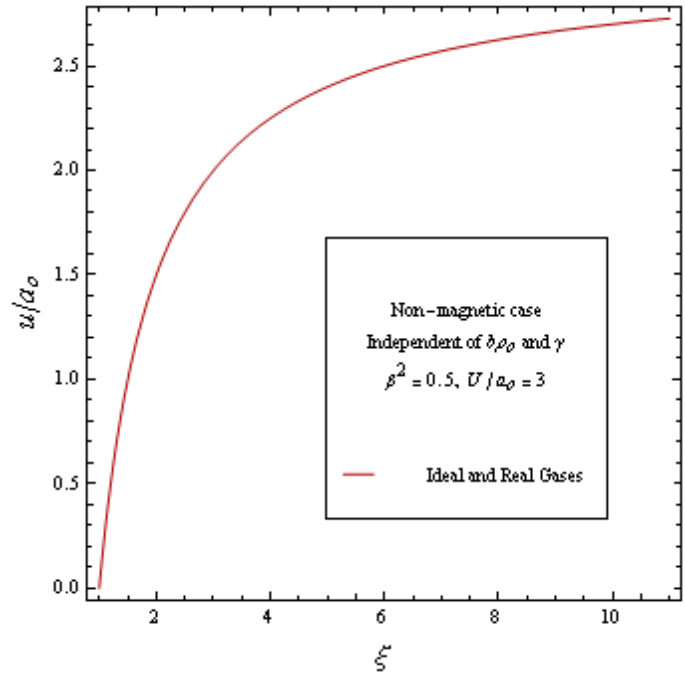

**Fig. 6** Variations of $p/p_o$, $a/a_o$ and $u/a_o$ with $\xi$ for different values of $b\rho_o$

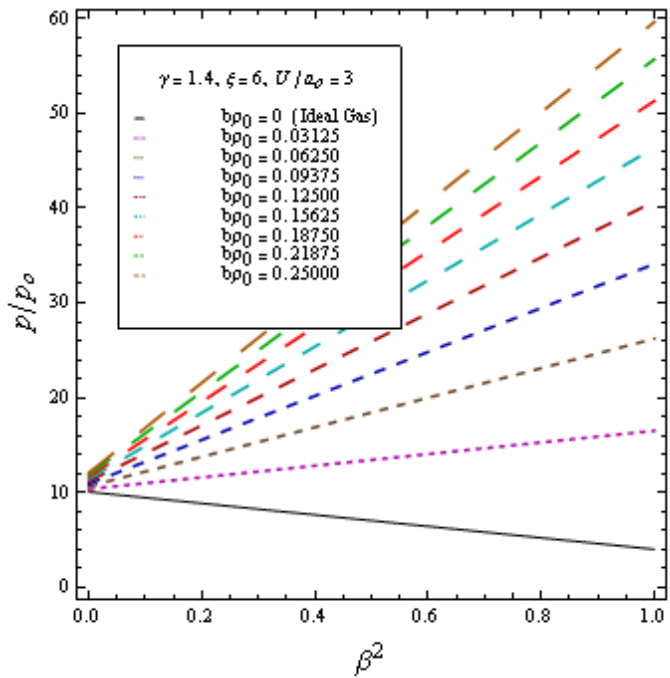

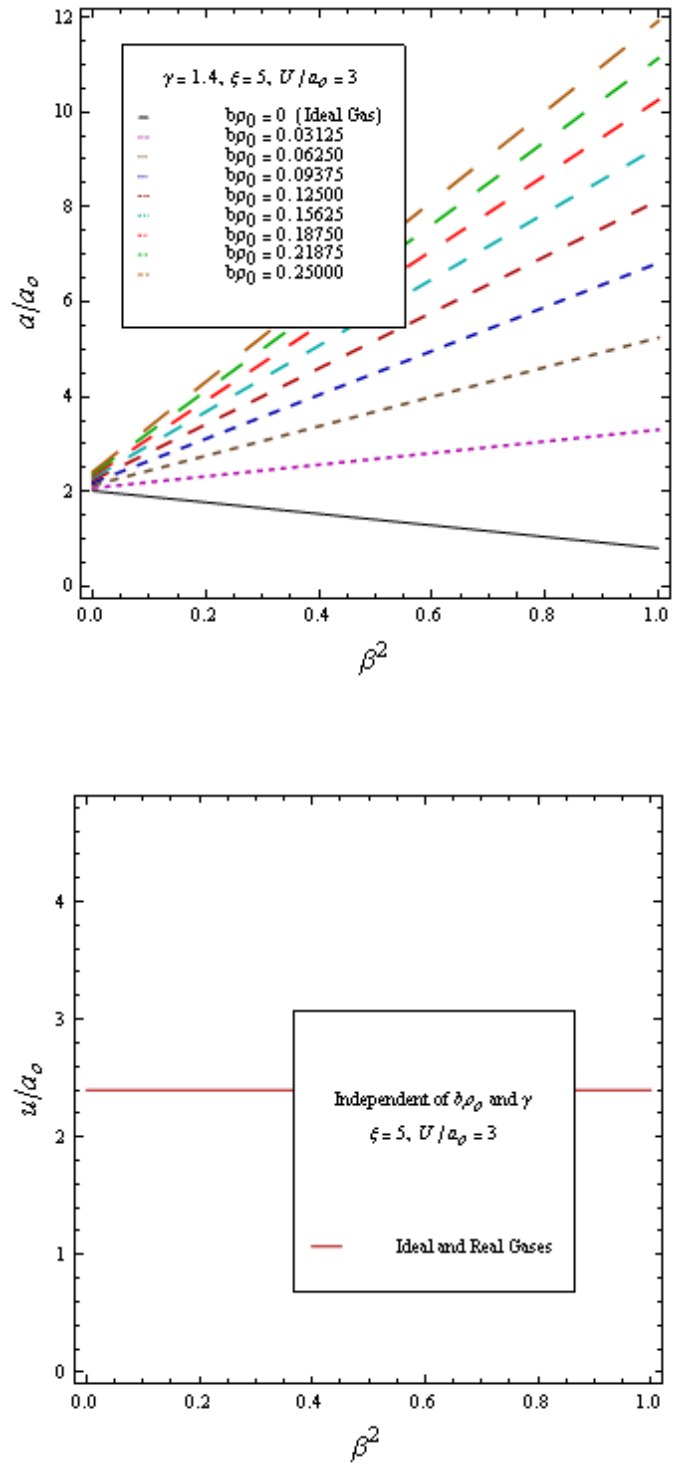

**Fig. 7** Variations of $p/p_o$, $a/a_o$ and $u/a_o$ with $\beta^2$ for different values of $b\rho_o$

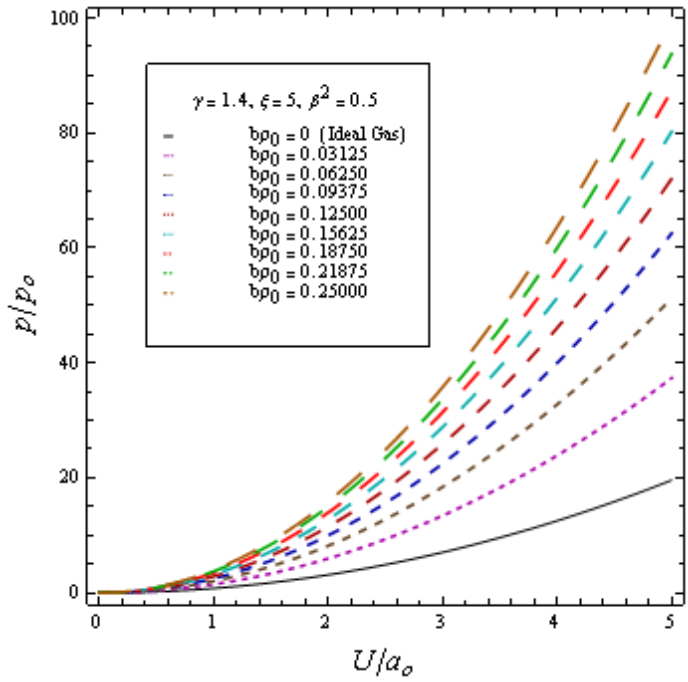

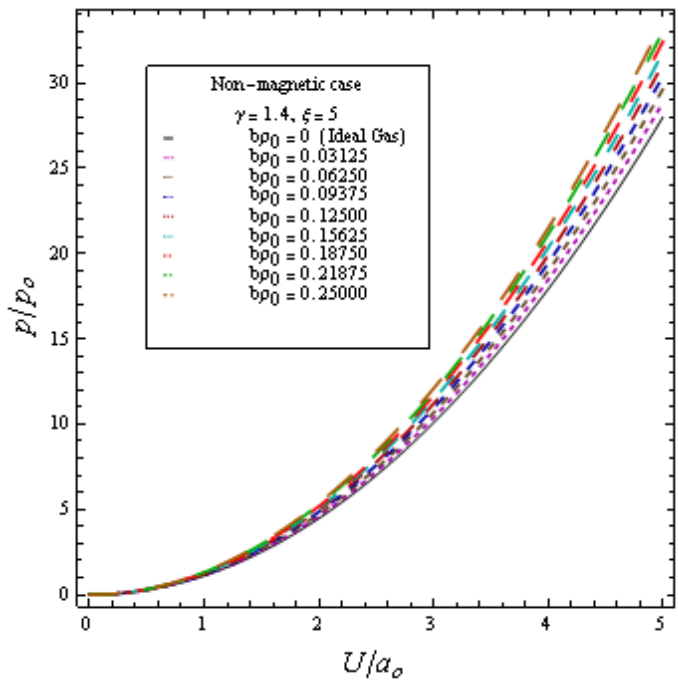

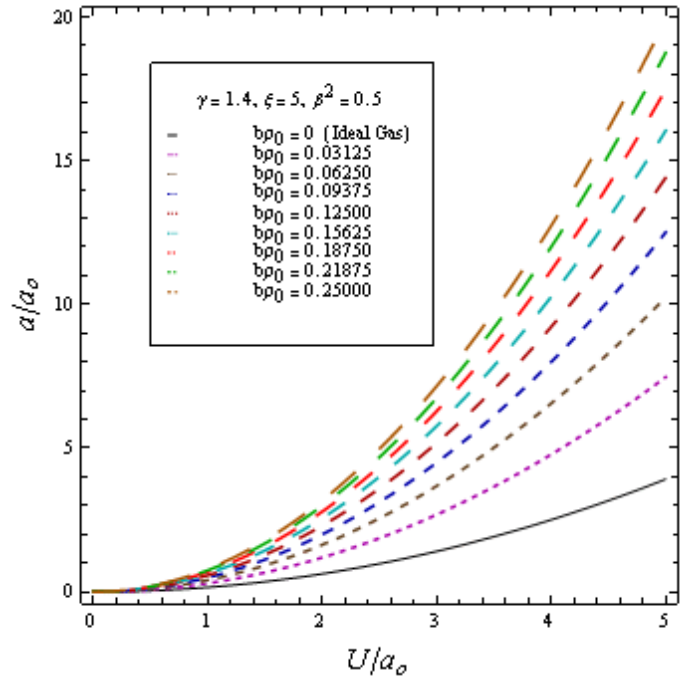

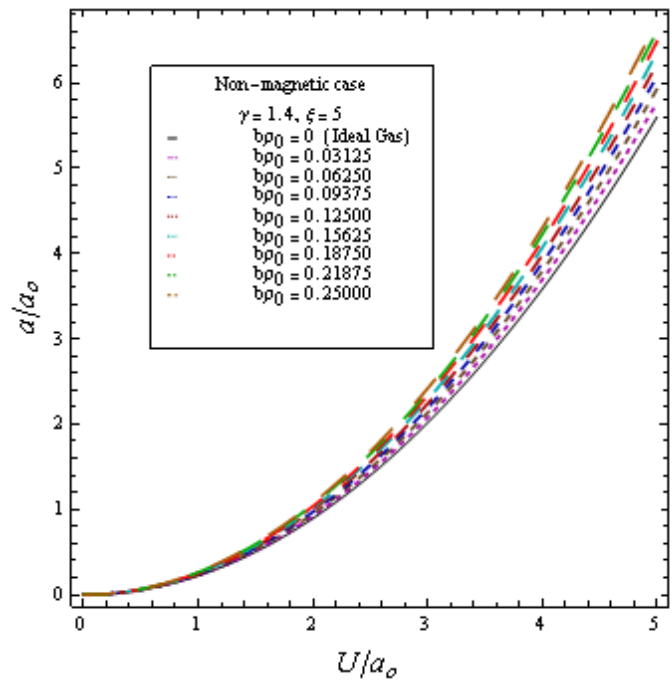

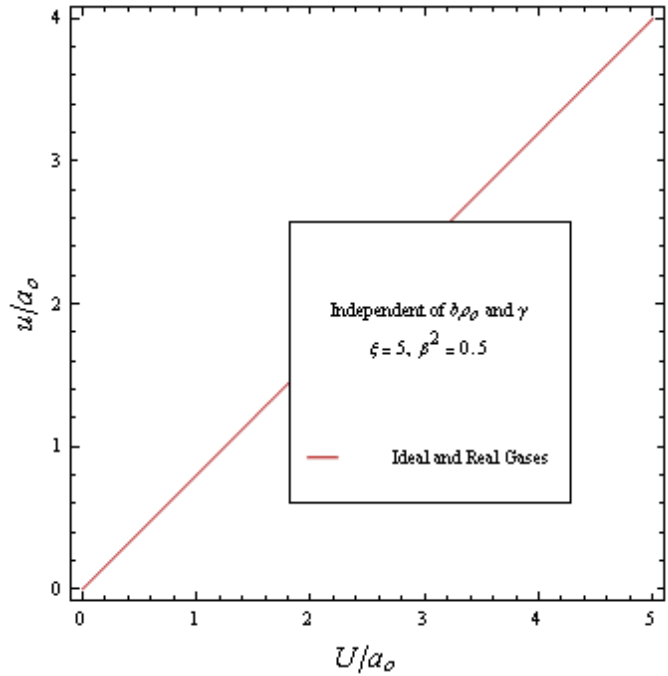

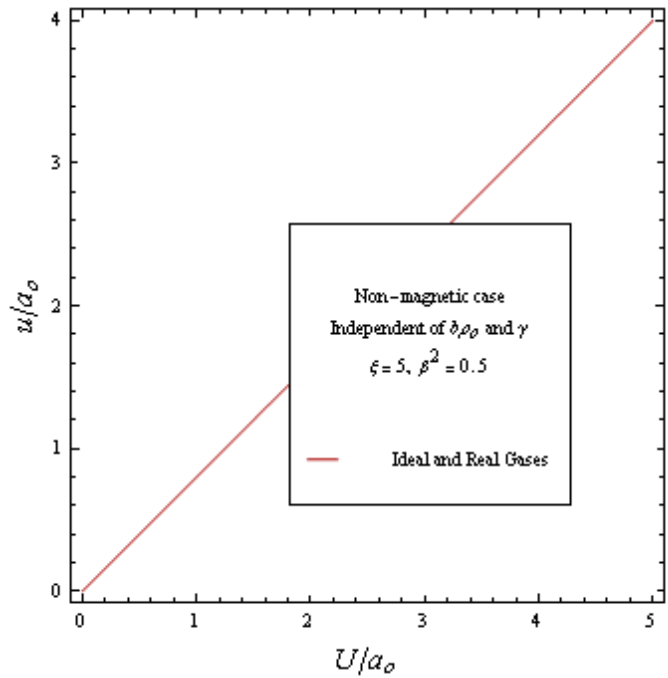

**Fig. 8** Variations of $p/p_o$, $a/a_o$ and $u/a_o$ with $U/a_o$ for different values of $b\rho_o$

4.2.2 Strong shock in strong magnetic field: The generalized MHD shock jump relations for strong shocks in strong magnetic field are given by Eq. (10). These relations i.e., pressure $p/p_o$ and particle velocity $u/a_o$ are dependent of $\xi$ a parameter characterizing the shock strength, strength of magnetic field $\beta^2$, adiabatic index $\gamma$, non-idealness parameter $b\rho_o$ and shock velocity $U/a_o$. The variations in the pressure, speed of sound and particle velocity with $\xi$ for $\gamma = 1.4$, $\beta^2 = 5$, $U/a_o = 7$ and different values of non-idealness parameter $b\rho_o$, are shown in Fig. 9. It is important to note that the pressure, speed of sound and particle velocity increase with increase in the shock strength. The speed of sound and particle velocity first increase and then become constant with the shock strength whereas the pressure increases continuously with the shock strength. It is also seen that an increase in the value of non-idealness parameter leads to an increase in the pressure, and speed of sound. It is also notable that the particle velocity does not depend on the non-idealness parameter, and adiabatic index. It is worth mentioning that the pressure and speed of sound first increase rapidly with the shock strength and achieve a maximum value and then start to decrease in the absence of magnetic field. The variation in particle velocity behind the shock is same in both the magnetic and non-magnetic cases. Obviously, in the presence of strong magnetic field the variations in the pressure and speed of sound are small as compared to the variations in the absence of magnetic field. Fig. 10 shows the variations in the pressure, speed of sound and particle velocity with $\beta^2$ for $\gamma = 1.4$, $\xi = 5$, $U/a_o = 7$ and different values of non-idealness parameter $b\rho_o$. It is important to note that the pressure and speed of sound except for the case of $b\rho_o = 0$, first slowly increase and then become constant with increase in the strength of magnetic field whereas for the case of $b\rho_o = 0$ the pressure and speed of sound first slowly decrease and then become constant with increase in the strength of magnetic field. It is also seen that an increase in the value of non-idealness parameter leads to an increase in the pressure and speed of sound. The particle velocity behind the shock remains unchanged with the strength of magnetic field. The variations in the pressure, speed of sound and particle velocity with $U/a_o$ for $\gamma = 1.4$, $\xi = 5$, $\beta^2 = 5$ and different values of non-idealness parameter $b\rho_o$, are shown in Fig. 11. It is important to note that the pressure, speed of sound and particle velocity increase with increase in the shock velocity. It is also seen that an increase in the value of non-idealness parameter leads to an increase in the pressure, and particle velocity. It is notable that the particle velocity does not depend on the non-idealness parameter, and adiabatic index but it increases linearly with the shock velocity for both the magnetic and non-magnetic cases. It is worth mentioning that the pressure, and speed of sound increase for $b\rho_o = 0$ and 0.03125 with increase in the shock velocity, whereas for other values of non-idealness parameter the pressure, and particle velocity decrease with the shock velocity. Very large variations in the pressure and speed of sound are noticeable with the shock velocity in the absence of magnetic field as compared to the variations in the presence of strong magnetic field. Obviously, the trends of variations of flow quantities behind the shock front in real gases are similar to that of behind the shock front in an ideal gas. It is notable that in the presence of strong magnetic field the pressure and speed of sound increase with the shock strength and shock velocity whereas the pressure and speed of sound decrease in the absence of magnetic field.

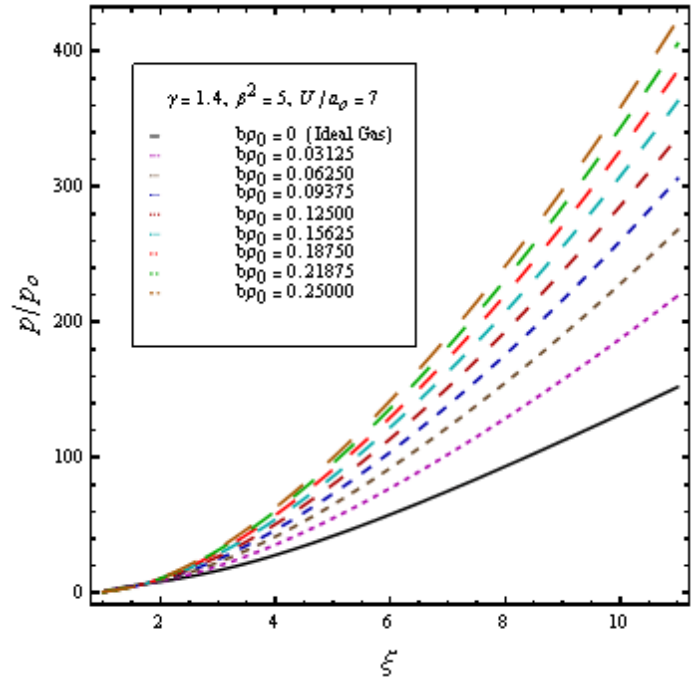

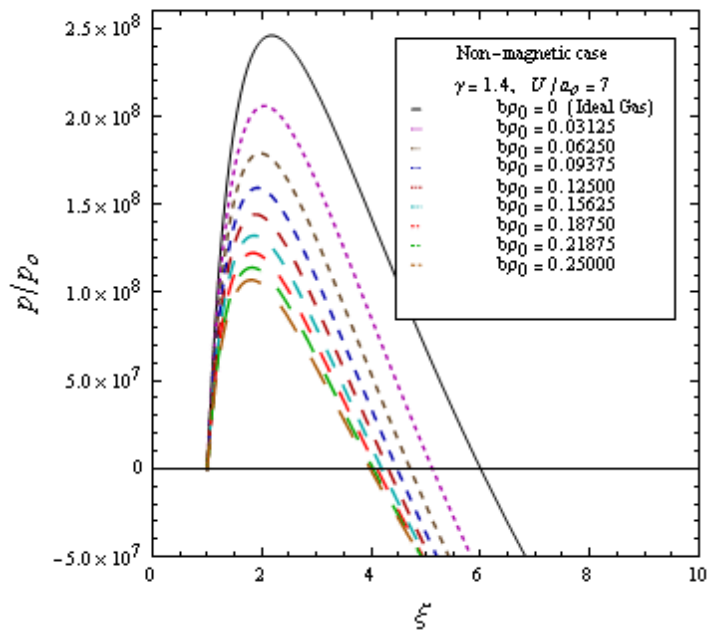

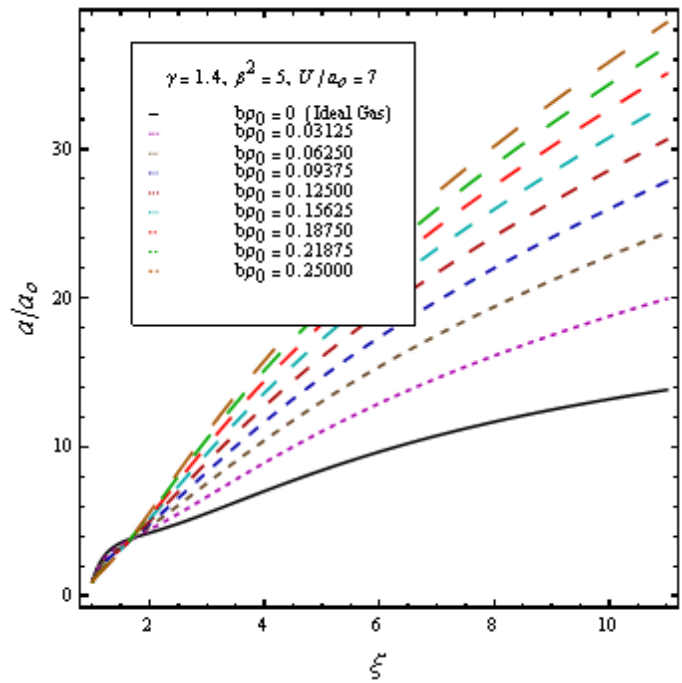

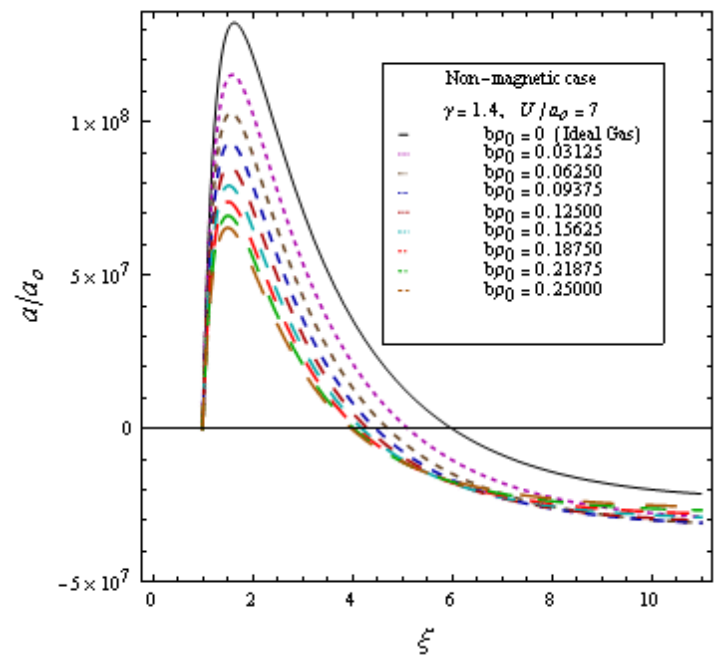

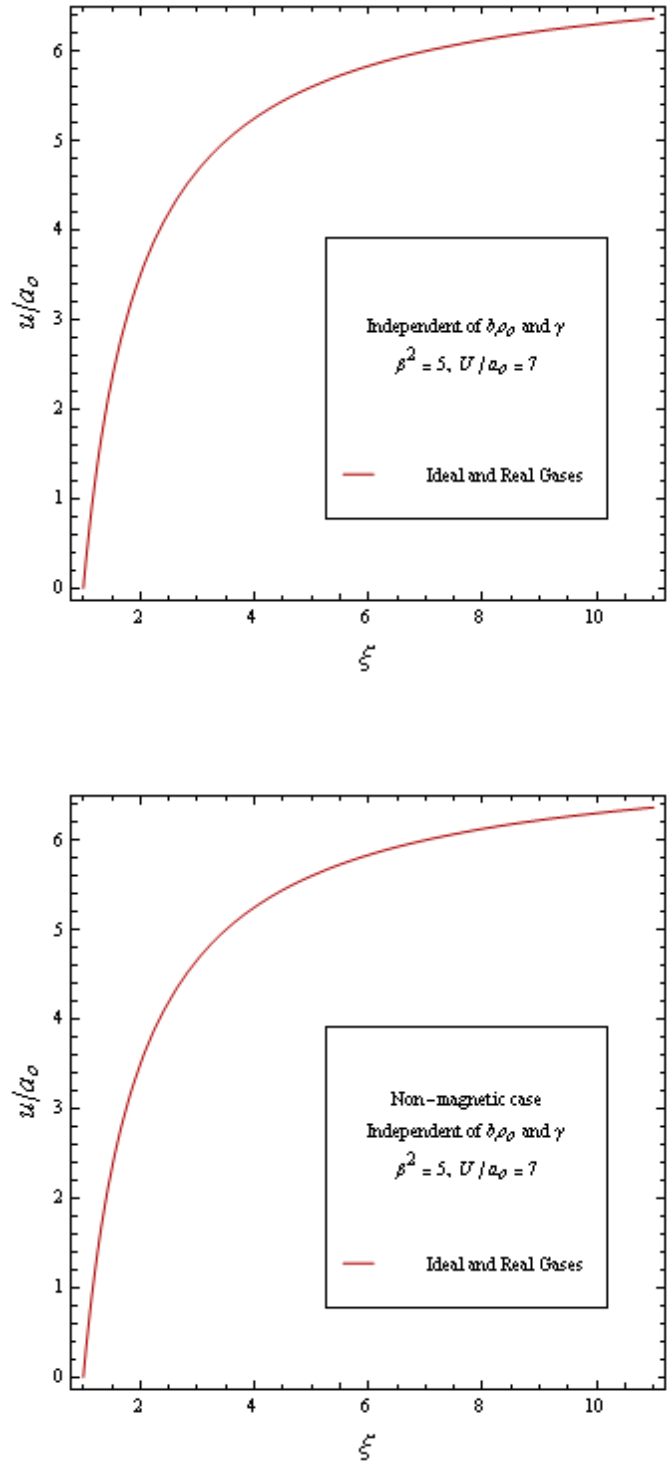

**Fig. 9** Variations of $p/p_o$, $a/a_o$ and $u/a_o$ with $\xi$ for different values of $b\rho_o$

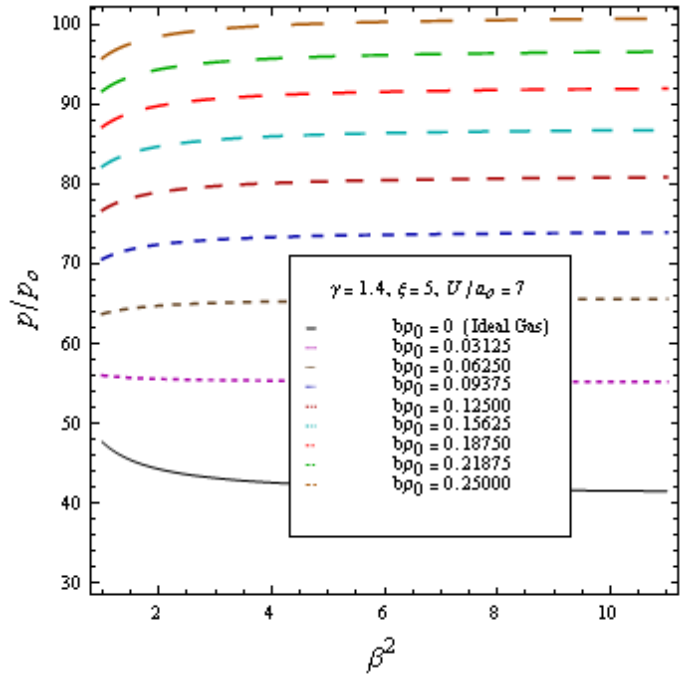

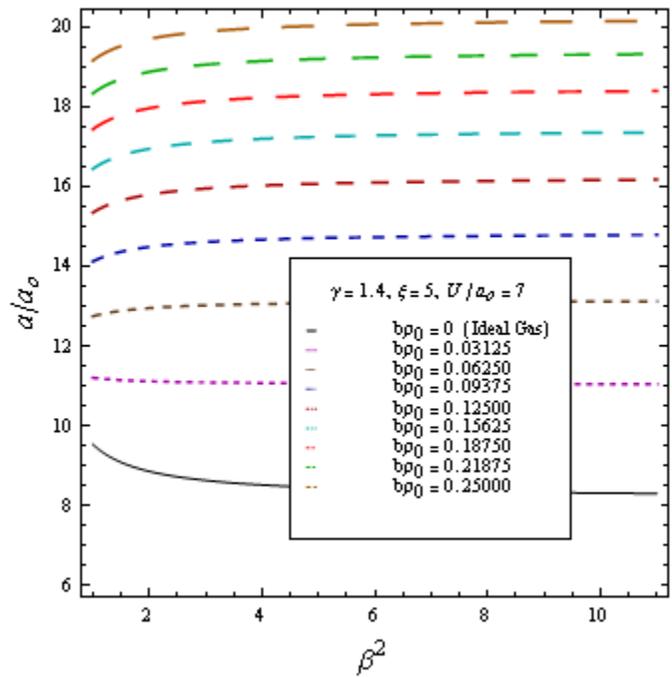

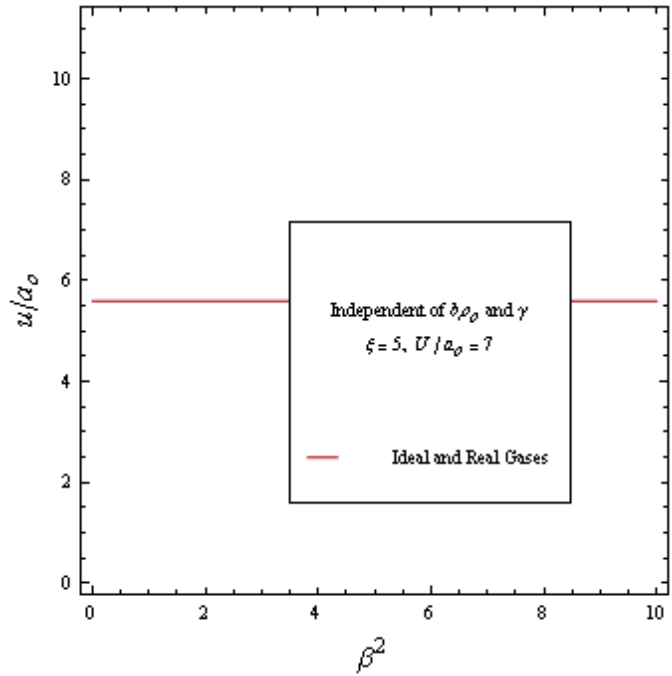

**Fig.10** Variations of $p/p_o$, $a/a_o$ and $u/a_o$ with $\beta^2$ for different values of $b\rho_o$

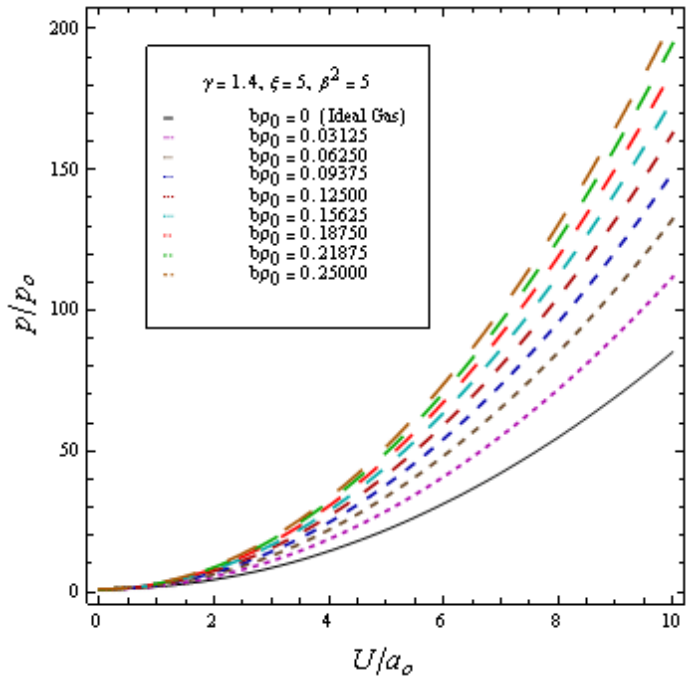

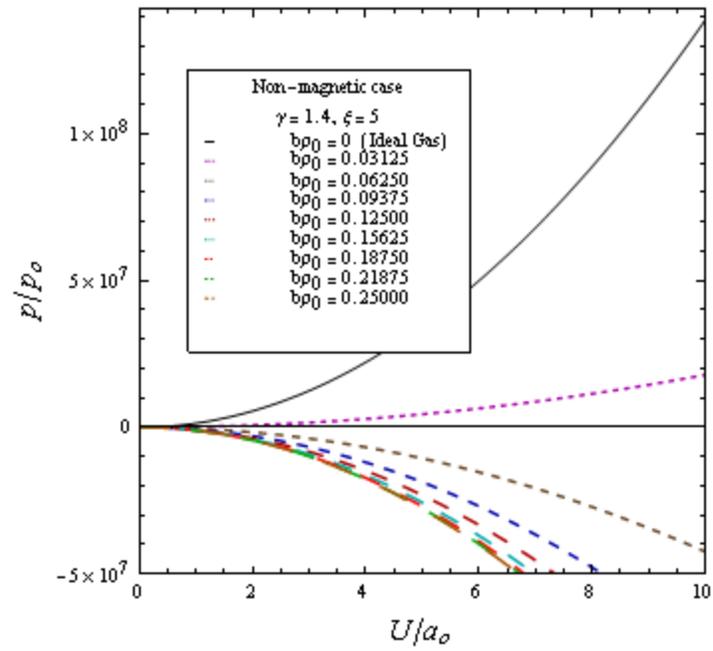

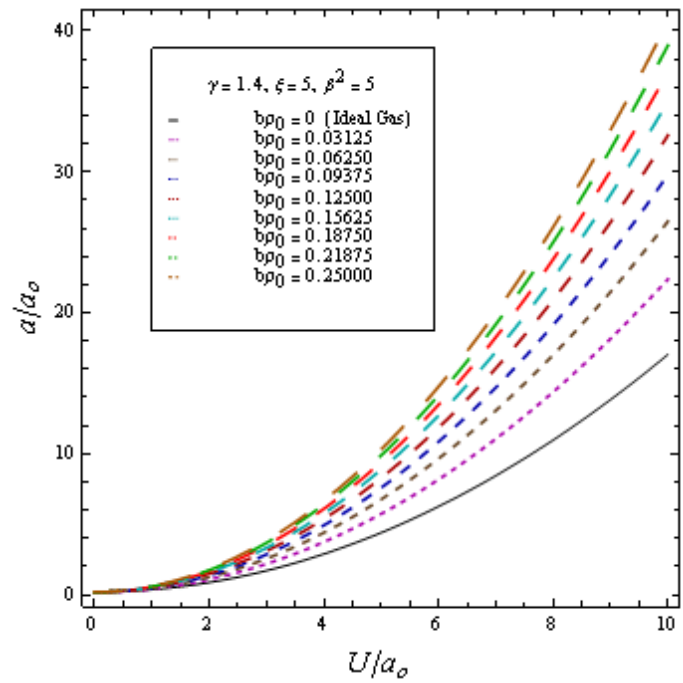

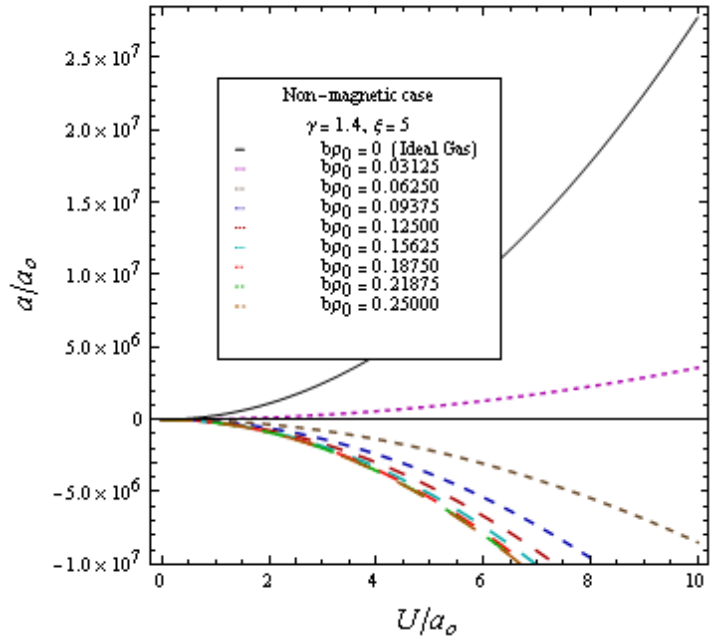

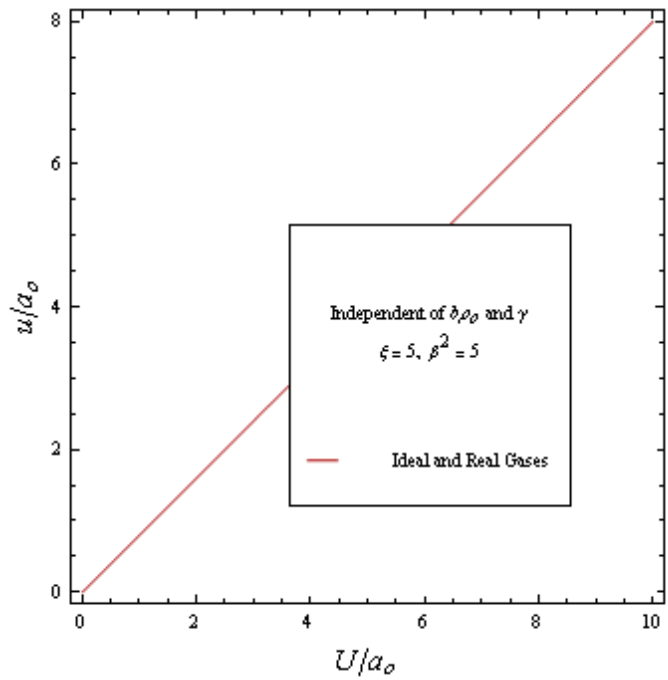

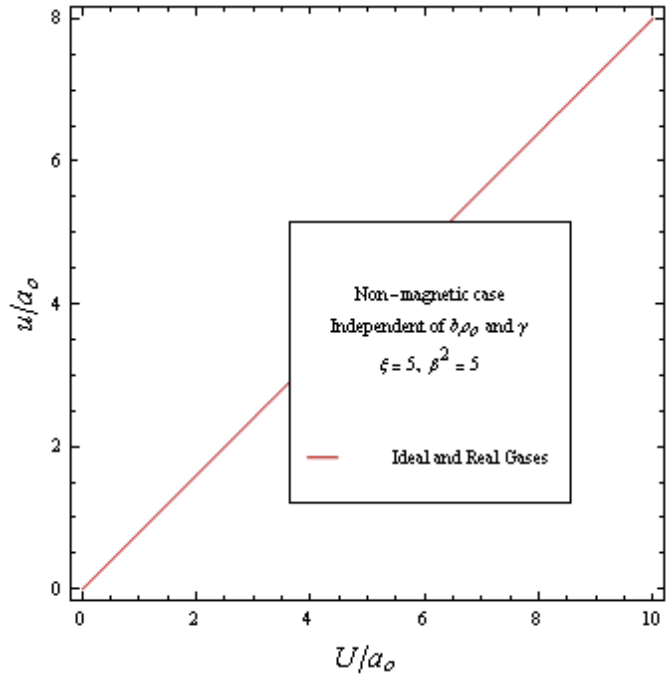

**Fig.11** Variations of $p/p_o$, $a/a_o$ and $u/a_o$ with $U/a_o$ for different values of $b\rho_o$

Applications: The handy forms of the generalized MHD shock jump relations derived in the present paper may be used directly to a wide variety of astrophysical problems, aerodynamics, gas turbines, internal combustion engines, fusion reactors, and a host of other computational fluid dynamics applications and also in many other fields including oceanography, and atmospheric sciences. The following are few astrophysical examples where the role of magnetic fields is important. The presence of sunspots in the photosphere, and structures such as filaments, prominences, and flares in the solar corona, demonstrate the key role of the magnetic fields. In fact, the very existence of the hot corona is now interpreted as due to heating by MHD effects. It is thought that most of the magnetic activity of the Sun is driven by the combination of rotation and turbulent flows in the convection zone. Both the processes that produce sunspots, and the large-scale magnetic field of the Sun, are very active areas of research. The key role of magnetic fields is also observed in the interstellar medium of galaxies. The observation of polarized synchrotron emission from the interstellar medium of the Milky Way and other galaxies, produced by relativistic electrons spiraling around magnetic field lines, is direct proof of the presence of such magnetic fields. Thus, interpretation of the dynamics of the interstellar medium requires an understanding of highly compressible MHD turbulence.

## 5. Conclusions and future plans

The analysis developed in the present paper is intended to complete the general studies on the MHD shock waves in real fluids by giving, for the first time, full analytic expressions of the generalized MHD shock jump relations explicitly in terms of the upstream parameters only, for weak as well as strong shock waves. This paper describes an investigation into the nature of MHD shock waves in real gases. The useful forms of the generalized MHD shock jump relations for weak and strong shocks provide simpler and more accurate boundary conditions to determine the MHD shock structure and propagation of MHD shock waves in non-ideal gases. However, although not exact, these handy forms of the generalized MHD shock jump relations have two advantages, (i) all shock relations are explicitly written in terms of the upstream parameters only and, (ii) the generalized forms of MHD shock relations are similar to the well-known MHD shock relations for an ideal gas (*see* Appendix). An important direct application of generalized shock jump relations is the study of propagation of shock waves produced due to supernova explosions, etc. The shock jump relations derived for SSWMF and SSSMF can be interpreted and applied in certain problems of steady supersonic flow in interstellar media.

The following conclusions may be drawn from the findings of the current analysis:

1. It is found that the shock velocity, pressure and particle velocity just behind the MHD shock wave in non-ideal gas flow increase with the strength of magnetic field, and non-idealness parameter. A large increase in the flow quantities behind shock is seen in the presence of magnetic field as compared to the absence of magnetic field. Thus, in the presence of magnetic field the shock propagates more rapidly in real gases comparatively in an ideal gas atmosphere. It is interesting to note that the rate of rise in the flow variables increases with the strength of magnetic field.

2. In case of WSWMF, it is worth mentioning that the properties of the weak shock waves are not affected by the presence of weak magnetic fields. The shock velocity, pressure and particle velocity decrease with the non-idealness parameter.

3. In case of WSSMF, it is notable that the shock velocity and particle velocity increase with the strength of magnetic field. The pressure except for $b\rho_o = 0$ decreases with the strength of magnetic field whereas for $b\rho_o = 0$ (Ideal gas) it remains constant. The shock velocity and pressure decrease with increase in the value of non-idealness parameter. The particle velocity is independent of the non-idealness parameter and adiabatic index however; it increases with the strength of magnetic field.

4. In case of SSWMF, it is important to note that the pressure and speed of sound increase (except for $b\rho_o = 0$) with increase in the strength of magnetic field, whereas the pressure and speed of sound decrease for $b\rho_o = 0$. The pressure and speed of sound increase with non-idealness parameter. The particle velocity does not depend on the non-idealness parameter and adiabatic index however it remains unchanged with the strength of magnetic field. It is important to note that the pressure, speed of sound and particle velocity increase with increase in the shock velocity.

5. In case of SSSMF, it is worth mentioning that the pressure, speed of sound, and particle velocity increase with increase in the shock velocity. The pressure and particle velocity increase with the non-idealness parameter but remain unchanged with the strength of magnetic field. The particle velocity does not depend on the non-idealness parameter and adiabatic index; however it remains unchanged with the strength of magnetic field. Very large variations in pressure and speed of sound are noticeable with the shock velocity in the absence of magnetic field as compared to the presence of strong magnetic field. Obviously, the trends of variations of flow quantities behind the shock front in real gases are similar to that of behind the shock front in an ideal gas. It is notable that the pressure and speed of sound in presence of strong magnetic field increase with the strength of shock and shock velocity whereas in the absence of magnetic field the pressure and speed of sound decrease with the strength of shock and shock velocity.

In general, the trends of variations of the flow quantities behind the shock wave in real gases are similar to that of behind the shock wave in an ideal gas. It is also desirable to mention that the effect of non-idealness parameter, generally, does not change the trends of variations of flow variables behind the weak and strong shock waves. In my view, the MHD shock waves problems can be generalized by making use of the generalized MHD shock jump relations obtained in the present paper and the generalized results can be easily reduced to the case of MHD shocks in an ideal gas atmosphere. Obviously, the generalized results will be closer to the actual situations. Thus, the generalized MHD shock jump relations have a key role in the research of shock waves in real gases, air, stars, interstellar medium, etc. The aim of this paper is to contribute to the understanding of how shock waves behave in the magnetized environments. Future work in this area is to obtain the analytic solutions for propagation of MHD shock waves in real gases. Further, this work may be extended to the astrophysical applications of MHD shock waves considering the effects of self-gravitation and rotation of fluids, and thus the generalized study of the MHD shock waves will automatically include the previous work.

Since astrophysical objects are normally not amenable to the experimental studies, astrophysicists seek some understanding by simulating them by computer simulations. With the advancement of computer technology and numerical algorithms, the complex astrophysical phenomena such as supernova explosions, accretion of material onto a star, stellar pulsations, or the granular pattern of solar convection are now accessible to simulation almost as if they were approachable via experiments in the laboratory. Many students and researchers of astrophysics are confronted with computer simulation results, or the prospect of executing simulation calculations, or even the writing of a simulation code. The present paper should provide some help in such circumstances. Moreover, it might also prove to be a valuable reference for the astrophysicists and fluid dynamicists.

**Acknowledgement**: I acknowledge the support and encouragement of my wife, Nidhi.

**Appendix: MHD shock jump relations for ideal gas (Whitham, 1958; Anand, 2000)**

$$\rho = \rho_o \xi, \quad H = H_o \xi, \quad u = U(\xi-1)/\xi,$$

$$U^2 = \frac{2\xi}{(\gamma+1)-(\gamma-1)\xi}\left[a_o^2 + b_o^2\left\{\left(1-\frac{\gamma}{2}\right)\xi + \frac{\gamma}{2}\right\}\right]$$

$$p = p_o + \frac{2\rho_o(\xi-1)}{(\gamma+1)-(\gamma-1)\xi}\left[a_o^2 + \frac{b_o^2}{4}\left\{(\gamma-1)(\xi-1)^2\right\}\right]$$

where $a_o = \sqrt{\gamma p_o/\rho_o}$ and $b_o = \sqrt{\mu H_o^2/\rho_o}$

Jump relations for WSWMF

$$\rho = \rho_o(1+\varepsilon), \quad H = H_o(1+\varepsilon), \quad u = \varepsilon a_o, \quad U = [1+(1+\gamma)\varepsilon/4]a_o, \quad p = p_o[1+\gamma\varepsilon]$$

Jump relations for WSSMF

$$\rho = \rho_o(1+\varepsilon), \quad H = H_o(1+\varepsilon), \quad u = \varepsilon b_o, \quad U = [1+3\varepsilon/4]b_o, \quad p = p_o[1+\gamma\varepsilon]$$

Jump relations for SSWMF

$$\rho = \rho_o \xi, \quad H = H_o \xi, \quad u = U(\xi-1)/\xi, \quad p/p_o = 1 + \delta\{\chi' a_o^2 + A' b_o^2\}U^2/a_o^4$$

where $\chi' = \dfrac{\gamma(\xi-1)}{\xi}$, $A' = \dfrac{\gamma(\xi-1)}{4\xi}\left[(\gamma-1)(\xi-1)^2 - 2\{(2-\gamma)\xi+\gamma\}\right]$

Jump relations for SSSMF

$$\rho = \rho_o \xi, \quad H = H_o \xi, \quad u = U(\xi-1)/\xi, \quad p/p_o = 1 + \chi\{b_o^2 + A a_o^2\}U^2/a_o^2 b_o^2$$

where $\chi = \dfrac{\gamma(\gamma-1)(\xi-1)^3}{2\xi\{(2-\gamma)\xi+\gamma\}}$, $A = \dfrac{4}{(\gamma-1)(\xi-1)^2} - \dfrac{2}{(2-\gamma)\xi+\gamma}$

**Jump relations for magnetohydrodynamic shock waves in non-ideal gas flow          R. K. Anand**
**Accepted for publication in Astrophysics and Space Science**Del Zanna, L., Amato, E., Bucciantini, N.: Astron. Astrophys. **421**, 1063 (2004)
Falgarone, E., Pety, J., Phillips, T. G.: Astrophys. J. **555**, 178 (2001)
FerrizMas, A. and MorenoInsertis, F.: Phys. Fluids A **4**, 2700 (1992)
Friedrichs, K. O.: Non-linear wave motion in magnetohydrodynamics, unpublished Los Alamos    Report (1955)
Fukumura, K., Tsuruta, S.: Astrophys. J. **611**,964(2004).
Genot, V.: Astrophys. Space Sci. Trans., **6**, 31 (2009)
Hartmann, L.: Accretion Processes in Star Formation, Cambridge University Press (1998)
Hennebelle, P.: Astron. Astrophys. **397**, 381 (2003)
Hoffmann, F. de., Teller, E.: Phys. Rev. **80(4)**, 692 (1950).
Kawachi, T., Hanawa, T.: PASJ, **50**, 577 (1998)
Klessen, R. S., Burkert, A.: ApJS, **128**, 287 (2000)
Koldoba, A. V., Ustyugova, G. V., Romanova, M.M. and Lovelace, R.V.E.: Mon. Not. R. Astron. Soc. **388**, 357 (2008)
Landau, L. D., Lifshitz, E. M.: Course of Theoretical Physics. Statistical Physics, vol. 5. Pergamon Press, Oxford (1958)
Lundquist, S.: Arkiv fur Physik **5**, 15 (1952)
Ostriker, E. C., Stone, J. M., Gammie, C. F.: Astrophys. J. **546**, 980 (2001)
Ouyed, R., Pudritz, R. E.: Astrophys. J. **419**, 255 (1993)
Philip, R., Shimshon, F.: Phys. Fluids **19**, 1889 (1976)
Pogorelov, N. V. and Matsuda, T.: Astron. Astrophys. **354**, 697 (2000)
Porter, D. H., Pouquet, A., Woodward, P. R.: Phys. Fluid, **6**, 2133 (1994)
Roberts, P. H. and Wu, C. C.: Physics Letters A213, 59 (1996)
Roberts, P. H. and Wu, C. C.: The shock wave theory of sonoluminescence. In shock focusing effect in medical science and sonoluminescence, edited by R C Srivastava, D Leutloff, K Takayama and H Gronig, Springer-Verlag (2003)
Sakurai, A.: J. Phys. Soc. Jpn. **8**, 662 (1953)
Sakurai, A.: J. Phys. Soc. Jpn. **9**, 256 (1954)
Sakurai, A.: J. Phys. Soc. Japan, **17**, 1663 (1962)
Sedov, L. I.: Similarity and Dimensional Methods in Mechanics, Chap. 4. Academic Press, New York (1959)
Shadmehri, M.:Mon. Not. R. Astron. Soc. **356**, 1429 (2005)
Singh, M., Singh, L. P., Husain, A.: Chin. Phys. Lett. **28**(9), 094701 (2011)
Singh, M.: Astrophys Space Sci (2012) doi: 10.1007/s10509-012-1262-8
Summer, D.: Astron.  Astrophys. **45**, 151 (1975)
Takahashi, M., Rillet, D., Fukumura, K., Tsuruta, S.: Astrophys. J. **572**, 950 (2002)
Tanaka T., Washimi H.: Science, **296**, 321 (2002)
Tilley, D. A., Pudritz, R. E.: Astrophys. J. **593**, 426 (2003)
van der Swaluw, E., Achterberg, A., Gallant, Y. A., Toth, G.: Astron. Astrophys. **380**, 309 (2001)
Whang, Y. C.: JGR., **89**, 7367 (1984)
Whitham, G. B.: Fluid Mech. **4**, 337 (1958)
Whitham, G. B.: Linear and Nonlinear Waves, Chap. 8. Wiley, New York (1974)
Wu, C. C., Roberts, P. H.: Q. J. Mech. Appl. Math. **49**(4), 501 (1996)
Zel'dovich, Ya. B., Raizer, Yu. P.: Physics of Shock Waves and High Temperature Hydrodynamic Phenomena, Chap. 2. Academic Press, New York (1966)42